  \documentclass[twocol]{ametsoc}

\usepackage{comment}

\journal{mwr}

\bibpunct{(}{)}{;}{a}{}{,}

\title{On Barotropic Mechanisms of Uncertainty Propagation \\in Estimation of Drake Passage Transport}

    \authors{Alexander G. Kalmikov\correspondingauthor{Alexander G. Kalmikov, 
     Department of Earth, Atmospheric, and Planetary Sciences, 
     Massachusetts Institute of Technology, 
     77 Massachusetts Ave, Cambridge, MA 02139.}
 and Patrick Heimbach\thanks{Current affiliation: 
      Institute for Computational Engineering and Sciences and Jackson School of      
      Geosciences, 
      The University of Texas at Austin,
      01 East 24th Street, POB 4.232, Austin, TX 78712.
    }}

     \affiliation{Department of Earth, Atmospheric, and Planetary Sciences,\\ 
     Massachusetts Institute of Technology, 
     Cambridge, Massachusetts}

\email{kalex@alum.mit.edu}

\abstract{
Uncertainty in estimation of Drake Passage transport is analyzed 
in a Hessian-based uncertainty quantification (UQ) framework.
The approach extends the adjoint-based ocean state estimation method to provide formal error bounds functionality. 
Mechanisms of uncertainty propagation in an idealized barotropic model of the Antarctic Circumpolar Current are identified by analysis of Hessian and Jacobian derivative operators, generated via algorithmic differentiation (AD) of the MIT ocean general circulation model (MITgcm). 
Inverse and forward uncertainty propagation mechanisms are identified, projecting uncertainty between observation, control and state variable domains. Time resolving analysis of uncertainty propagation captures the dynamics of uncertainty evolution and reveals transient and stationary uncertainty regimes. 
The UQ system resolves also the dynamical coupling of uncertainty across different physical fields, as represented by the off-diagonal posterior covariance structure. 
The spatial patterns of posterior uncertainty reduction and their temporal evolution are explained in terms of barotropic ocean dynamics. 
Global uncertainty teleconnection mechanisms are associated with barotropic wave propagation. 
Uncertainty coupling via data assimilation is demonstrated to dominate the reduction of Drake Passage transport uncertainty, highlighting the importance of correlation between different oceanic variables on the large scale.  
} 

\begin{document}

\maketitle

%
\section{Introduction}

\begin{figure}[h]
 \centerline{
  \includegraphics[width=19pc]{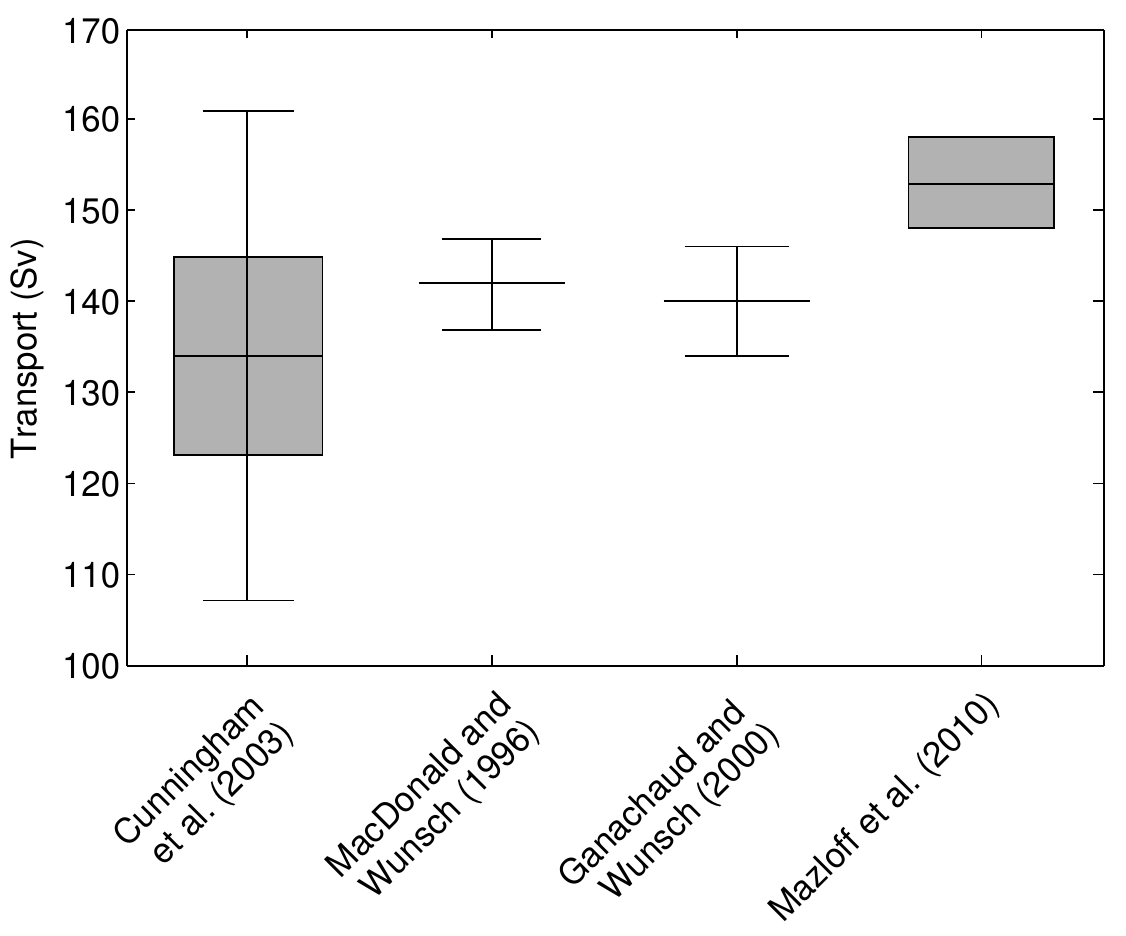}
 }
 \caption{Drake Passage transport estimates from ISOS and WOCE hydrographic measurement campaigns. Estimated transport uncertainties are shown with error bars, variability is shown with shaded boxes.} 
  \label{obs-Transport} 
\end{figure}

The strength of the zonal (west-to-east) volume transport carried by the Antarctic Circumpolar Current (ACC) -- the largest current system in the world ocean, is a key climatological index and a standard diagnostic of numerical global ocean models \citep{Olbers2004,Farneti2015}. 
ACC transport is a central component of global ocean circulation, connecting the three major ocean basins and regulating the global heat, freshwater and greenhouse gas exchange. It acts as a barrier to meridional transport, isolating the Antarctic continent from warmer ocean water and explaining its glacial ice \citep{Rintoul2001}.
Traditionally monitored at Drake Passage --  the narrowest constriction along the ACC's path \citep{Meredith2011} -- the volume of its circumpolar transport has been a subject of debate of many observational, theoretical and modeling studies. 
Estimates in the literature range from well under 100~Sv to over 200~Sv  
\citep[Table 1]{Peterson1988}. 
Agreement on the exact value is complicated by a substantial natural variability of the transport as well as uncertainty in the estimation procedure. 
Detailed error analysis of the International Southern Ocean Studies (ISOS) experiment hydrographic transport estimate of \citet{Whitworth1985} by \citet{Cunningham2003} explicitly separates these two uncertainty factors, reporting for the 1975 to 2000 period a steady average transport of 134~Sv with temporal standard deviation 11.2~Sv and estimation uncertainty of 15 to 27~Sv. 
Modern inverse model-based transport estimates, however, assimilating the World Ocean Circulation Experiment (WOCE) data, do not separate uncertainty and variability. In global ocean circulation steady-state estimation with hydrographic inverse box models \citep{GanachaudWunsch2000}, rigorous uncertainty bounds for the steady circulation estimate conflate uncertainty due to variability and estimation error into a single number, aliasing variability in one-time hydrographic sections \citep{Ganachaud2003c}. The reported inverse box models' transport uncertainties are $142 \pm 5$~Sv \citep{Macdonald1996} and $140 \pm 6$~Sv \citep{GanachaudWunsch2000} in 1990. These can be compared (Figure~\ref{obs-Transport}) with the pure temporal variability-based uncertainty estimate $153 \pm 5$~Sv in the Southern Ocean State Estimation (SOSE) estimate for 2005--2006 \citep{Mazloff_etal2010}, but which omits a formal estimation uncertainty analysis because of the underlying computational challenge.  
It is this gap that we address in this paper using the framework developed in \citet{Kalmikov2013}.

The need for practical uncertainty quantification (UQ) methods in modern ocean and climate modeling is a recognized outstanding challenge \citep{Wunsch2006}. 
Separating uncertainty arising from observation and model errors from naturally occurring variability is necessary for reconciling inconsistency of conflicting transport estimates, while focusing further research on uncertainty reduction. Since uncertainty evolves dynamically in ocean models, not distinguishing between variability and error statistics evolution has led to controversies in the data assimilation literature \citep{Lermusiaux1999a}. 
It has also inhibited the broader climate science communication \citep{Lehmann2014Nature}. Here we address this need by introducing a Hessian-based method developed for uncertainty quantification in ocean state estimation \citep{Kalmikov2014} and prototype its application to formal analysis of estimation uncertainty in barotropic transport through Drake Passage. To disentangle quantification of estimation uncertainty from natural variability in this analysis we focus on uncertainty dynamics in a steady state ACC model. By construction, variability is zero and the resolved evolution of transport uncertainty is only due to the propagation of uncertainties in the model.  

The Hessian UQ method \citep{Kalmikov2014} was developed to formally assess confidence bounds in adjoint method-based ocean state estimation systems such as ``Estimating the Circulation and Climate of the Oceans''
\citep[ECCO\footnote{\url{http://ecco-group.org}},][]{Wunsch2009}. 
Our method extends the  adjoint method \citep{Wunsch2007} to employ the second order geometry information of the least-squares misfit function by inverting its Hessian matrix to evaluate large error covariance matrices of the estimated ocean state. 
Closely related to the second-order adjoint method \citep{Wang1992,LeDimet2002}, our algorithmic differentiation (AD) machinery computes projections of second derivative information on vectors (i.e. matrix-free) to machine precision \citep{Griewank-Walther:2008}, but has the advantage of flexibility and speed thanks to automated derivative generation for different model configurations. Our AD method generates the full Hessian matrix directly, not by aggregation of first derivatives in a quasi-Newton process \citep{Veerse2000b,Gejadze2008b}  nor is limited to evaluation of only the linearized (Gauss-Newton) Hessian part, such as in the ROMS model \citep{Moore2011}, an advantage for nonlinear large residual estimation problems. The adjoint method, also known as variational data assimilation or 4DVAR \citep{COURTIER1993, Anderson1996}, estimates a dynamically consistent three-dimensional time evolving ocean state by optimal fit of an ocean general circulation model (GCM) with diverse ocean observations. Key advantage of the adjoint method, unlike  sequential data assimilation and reanalysis filtering approaches -- the dynamical consistency constraints avoid spurious imbalances of conserved physical quantities \citep{WunschHeimbach2013,Stammer2016}. The accompanying Hessian method was designed to preserve this dynamical consistency and eliminate numerical artifacts in the resolved uncertainty dynamics \citep[see discussion in][]{Kalmikov2014}. 

\begin{figure*}[t]
 \centerline{
  \includegraphics[width=28pc]{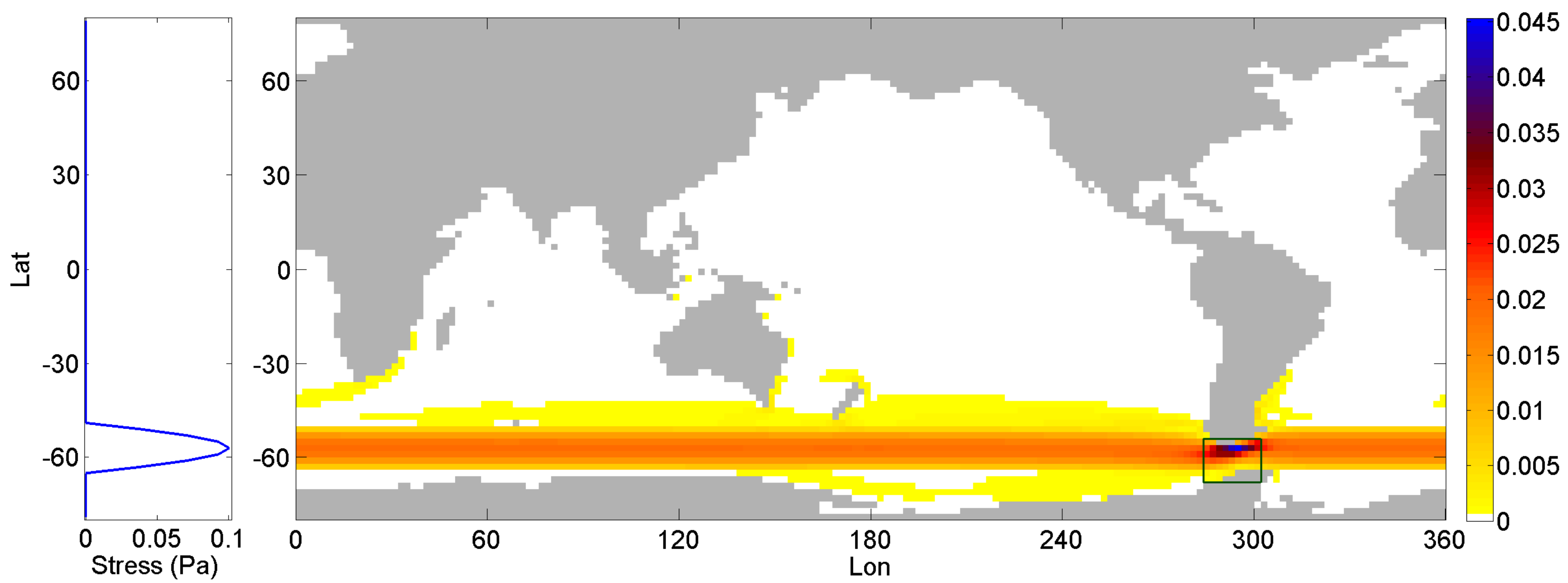}
 }
 \caption{Barotropic MITgcm configuration of ACC forced by idealized zonal wind stress profile (a). The steady state circumpolar circulation (b) is visualized by the magnitude of barotropic flow (color shades, m/s). The area of assimilated sea surface height observations at the Drake Passage is shown with black rectangle.} 
  \label{fig_configurationDomain}
\end{figure*}


The present paper extends the technical overview of the Hessian method
development of \citep{Kalmikov2014} to report the following innovative features and insights.  
The analysis of uncertainty propagation in the model is time-resolving, capturing the dynamics of uncertainty evolution and revealing transient and stationary uncertainty regimes. It quantifies the effects of assimilated observations' uncertainty on model solutions, reducing uncertainties of model controls and the target quantity of interest (here ACC transport). Inverse and forward uncertainty propagation mechanisms are contrasted demonstrating different time dependencies projecting the assimilated uncertainty between observations, controls and target quantities of interest.
Dynamical coupling of uncertainty between different physical fields is resolved by the off-diagonal posterior covariance terms and is shown to dominate uncertainty reduction. The application of the UQ method to Drake Passage transport reveals the dynamics of the spatio-temporal structure of the uncertainty of the estimated ocean state, highlighting uncertainty teleconnections and its adjoint-time evolution. 

The remainder of the paper is organized as follows: the barotropic model used to simulate Drake Passage transport and its state estimation framework are described in section 2; the Hessian UQ scheme is summarized in section 3; the resolved barotropic mechanisms of uncertainty propagation are analyzed in section 4. A summary and conclusions are offered in section 5. 


\section{Drake Passage transport estimation in idealized global model}
\subsection{Forward model of barotropic wind driven circumpolar transport}

Due to the lack of meridional boundaries which constrain oceanic flows in other basins, the dynamics of the ACC resembles 
qualitatively that of the atmospheric jet stream, \citep{Thompson2008} 
governed by geostrophic and thermal wind balances, while the Sverdrup balance does not hold.
Quantitatively, details of ACC dynamics are complex, involving multiple scales and superposition of barotropic and baroclinic mechanisms \citep{Marshall2003,OlbersLettmann2007,Nadeau2012}. 
Nonetheless, a widely held tenet in oceanography \citep{Cunningham2003} is that the variability of ACC transport through Drake Passage is dominated by barotropic, i.e. vertically integrated processes of flow response to varying wind forcing \citep{Whitworth1985,Hughes1999a,Weijer2005}. We focus on uncertainty dynamics on these scales and, for consistency with variability studies, concentrate on a barotropic model of ACC. 

Barotropic wind driven response of the circumpolar transport is modeled in an idealized global vertically integrated single layer configuration of the MITgcm\footnote{\url{http://mitgcm.org/}} \citep{Marshall1997a,Marshall1997}. The nonlinear shallow water momentum and continuity equations
\begin{equation} \label{eqShallowWater} 
\begin{array}{l}
 \displaystyle{\frac{{\partial u}}{{\partial t}} + uu_x  + vu_y  = fv - g\eta _x  + \frac{{\tau ^x }}{{\rho H}} - \frac{{ru}}{H} } \\ 
 \displaystyle{\frac{{\partial v}}{{\partial t}} + uv_x  + vv_y  =  - fu - g\eta _y  + \frac{{\tau ^y }}{{\rho H}} - \frac{{rv}}{H} }\\ 
 \displaystyle{\frac{{\partial \eta }}{{\partial t}} =  - \left( {u(H + \eta )} \right)_x  - \left( {v(H + \eta )} \right)_y  }\\ 
 \end{array} 
\end{equation}
are solved in a homogeneous 5000~m deep domain with realistic continental outlines (Figure \ref{fig_configurationDomain}b) and 
2 degrees horizontal resolution. 
Here $u$, $v$ are vertically integrated (barotropic) velocities, $\eta$ is the sea surface elevation, $H$ is the ocean depth, $f$ and $g$ are Coriolis and gravity parameters, and $\rho$ is density. 
$\tau ^x$ and $\tau ^y$ are components of the surface wind stress forcing the flow, here represented by an 
idealized zonally homogeneous constant in time zonal jet with sinusoidal meridional profile (Figure \ref{fig_configurationDomain}a). The balancing effects of bottom friction and form drag are parameterized with a linear bottom drag coefficient, assumed, following \citet{LoschWunsch2003}, $r=5 \times 10^{-3}$ m~s$^{-1}$. For 5000~m depth it translates to a spin down time scale $H/r=10^6$~s or about 12 days. 
The amplitude of wind stress is set to
$\tau ^x  = 10^{- 1}$~Pa 
to balance a typical zonal flow rate of $U=0.02~$m~s$^{-1}$,
i.e. the scale of the kinematic wind stress is 
$\tau ^x /\rho  = rU = 10^{ - 4}$~m$^2$~s$^{-2}$.
Rossby number of the simulated flow is 
${\rm{Ro = }}U{\rm{(}}Lf)^{ - 1}  = 2 \times 10^{ - 4} $, 
consistent with barotropic dynamics of Drake Passage transport.

\begin{figure}[t]
 \centerline{
  \includegraphics[width=19pc]{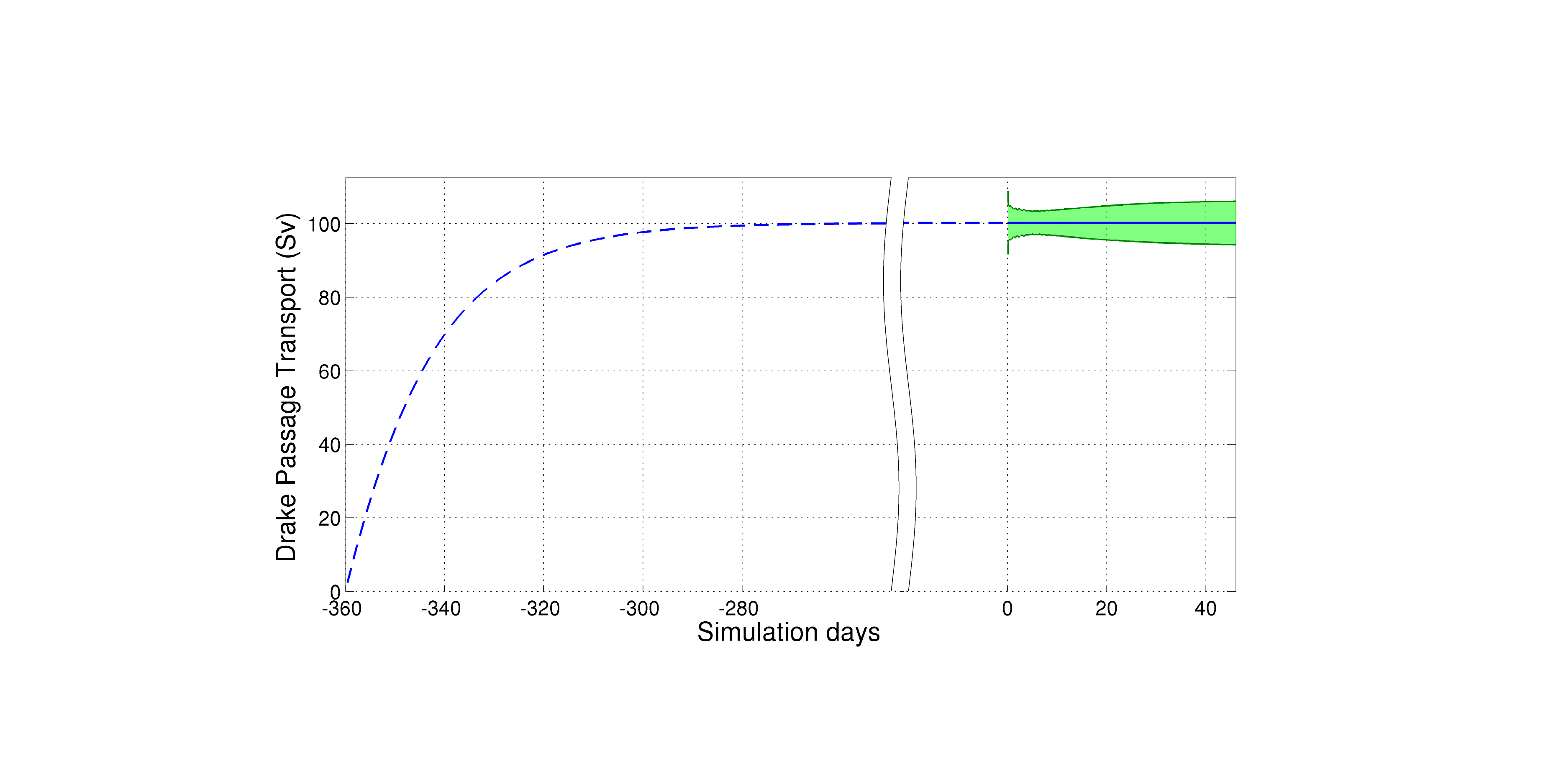}  
 }
 \caption{Time evolution of simulated Drake Passage transport and its uncertainty. The model is initialized from zero flow initial conditions and spun up for 360 days, growing asymptotically to a steady state. The uncertainty of the steady state flow evolves from the reinitialization time, when it is exlicitly specified in terms of the initial flow conditions, and exhibits a non-trivial temporal dynamics analysed and explained below. Time axis is shown excluding days from -250 to -20.} 
  \label{fig_TransportGrowth}
\end{figure}

Under steady westerly winds 
this system is governed, to leading order, by adjustment to a geostrophic forcing-friction equilibrium. The model is spun up from rest to the geostrophic steady state (Figure~\ref{fig_configurationDomain}b). In the adjustment process the flow is accelerated eastward by the wind and is deflected equatorward by the Coriolis force. This leads to a transient northward transport, redistributing the water mass across the jet and generating a northward gradient of sea surface elevation, which balances the Coriolis force in the equilibrium. 
The resulting geostrophic steady state flow is parallel to sea surface elevation contours and balanced by bottom friction. The circumpolar flow and sea surface height contours are aligned with the zonal wind forcing, except in the narrowing of  Drake Passage, where the contours are tighter and the flow is stronger, especially on the northern side of the Passage. The simulated barotropic flow results in a zonal volume transport of 100 Sv.

The meridionally integrated transport grows from zero at initial time and exponentially adjusts to the steady state equilibrium, Figure~\ref{fig_TransportGrowth}. The adjustment time scale of the transport is given by the frictional spin-down rate, which in turn is inversely proportional to the bottom friction coefficient and is linear in the depth of the ocean $T_{\rm friction}=H/r$. In deep basins or when the friction is weak, this is a slow time scale governing the dynamics of ACC evolution, leading to a linear ODE parameterization of the exponential transport dynamics \citep{WearnBaker1980,kalmikov2013barotropic}.
 The numerical simulation, however, reveals additional barotropic time scales modulating the circumpolar volume transport. On shorter time scale of few hours the flow is modulated by fast Kelvin waves propagating along continental coasts and the equatorial wave guide. Slower planetary Rossby waves propagate across the Pacific ocean on time scale of few days. Horizontal momentum diffusion stabilizes the numerical simulations, dissipating residual flow perturbations on time scale of a year \citep[][Section 5.1.4]{Kalmikov2013}.

\subsection{Inverse model configuration}

A state estimation problem is formulated for constraining the circumpolar volume transport by its indirect observations at the ocean surface. This is a canonical problem in ocean estimation, because measurement of deep oceanic flow is not feasible at sufficiently high spatial and temporal resolutions to characterize it directly. On the other hand, high-resolution remote sensing of the surface signature of the deep flow enables its indirect inference \citep{WunschStammer1998}, utilizing the dynamical coupling between barotropic transport and geostrophic sea surface elevation patterns. The ACC transport is estimated at the Drake Passage, where its spatial extent is most constrained and dynamical signals are stronger. 
The transport is meridionally-integrated across 
the Drake Passage at time $t$, constituting a time-dependent scalar target quantity of interest (QoI) $z(t)$. 

Synthetic surface elevation observations $\eta(t_a)$, generated in an identical twins setup \citep[Section 2.10]{Kalmikov2013} in the rectangular area over the Drake Passage ({Figure}~\ref{fig_configurationDomain}b), are assimilated in the barotropic model at assimilaton time $t_a$. 
Assimilation of surface observations’ uncertainty quantifies the extent of deep flow constraint by the barotropic dynamics linking it with its surface signature.  Analysis of the barotropic uncertainty evolution reveals the physical mechanisms of these links, quantifying the patterns and dynamics of barotropic uncertainty propagation.

\begin{table}[h!]
\centering
\caption{Schematic arrangement in the six-panel figures of the six two-dimensional control variables of the composite control vector $\bf{x}$.
\label{tab:tab1}}
\begin{tabular}{|l|l|}
\hline
\textbf{Boundary Conditions} & \textbf{Initial Conditions} \\
\hline
Zonal wind stress $\tau^x$ & Zonal flow $u_{t_0}$ \\
Meridional wind stress $\tau^y$ & Meridional flow $v_{t_0}$ \\
Bottom drag coefficient $r$ & Surface elevation $\eta_{t_0}$ \\
\hline
\end{tabular}
\end{table}  

The vector of control\footnote{For brevity, in lieu of linear control theoretic formalism of \citep{Wunsch2007} we adopt here a nonlinear inverse theoretic notation, where unknown model parameters, initial state conditions and boundary conditions controls are combined to form a generalized vector of estimated unknowns $\bf{x}$ controlling the model solution trajectory in the state space. See details in
\citep[Section 2.12]{Kalmikov2013} and
\citep[Section 3.1]{Nguyen2011}.} 
variables $\bf{x}$ in the present idealized inverse problem configuration is comprised of 6 two-dimensional fields of model initial and boundary conditions (Table 1), consisting of 80 by 180 scalar elements each. The length of the control vector is $O(10^{5})$ scalar elements. The Hessian matrix consists of $O(10^{10})$ elements, requiring 60 GB of uncompressed storage. In realistic global oceanographic inverse problems the dimensionality of the estimated state vector is several orders of magnitude higher, $O(10^{6})-O(10^{11})$ depending on time stepping and resolution \citep{Wunsch2007}, necessitating matrix-free and reduced order implementation of the algorithms as the only feasible approach. The developed matrix-free reduced Hessian UQ machinery \citep{Kalmikov2013} is designed to scale for high dimensional problems and has been demonstrated to compress the dimension of Hessian computations from the inverse problem state space size to the number of independent information-carrying observations \citep{Kalmikov2014}.

\section{Uncertainty Quantification method}

The Hessian UQ method \citep{Kalmikov2014} was developed for large-scale ocean state estimation in which a nonlinear model $\textbf{\textit{M}}(\bf{x})$ is fit in a least-squares sense to observations $\bf{y}$. The minimized quadratic misfit function
\begin{equation} \label{eqCost}
J_1({\bf{x}}) = \tfrac{1}{2} 
 (\textbf{\textit{M}}({\bf{x}}) - {\bf{y}})^T {\bf{R}}^{ - 1}  
 (\textbf{\textit{M}}({\bf{x}}) - {\bf{y}})
\end{equation}
 is given by a weighted inner product of model-data discrepancy vectors normalized by the covariance of observation and model error uncertainties $\bf{R}$. 
Large control error covariance matrices 
\begin{equation} \label{eqCov} 
\textbf{P}_{\bf{xx}}=\left\langle \bf(\hat{x}-x)(\hat{x}-x)^{\it T}  \right\rangle, 
\end{equation}
i.e. the dispersion of the estimate $\bf \hat{x}$ around the unknown true value $\textbf{x}$,
are evaluated by inverting the Hessian
\begin{equation} \label{eqHessian} 
{\bf{H}} ({\bf{x}}) \equiv \frac{{\partial ^2 J_1}}{{\partial {\bf{x}}\,\partial {\bf{x}}^T }}
\end{equation}
of model-data misfit using matrix-free numerical linear algebra algorithms. 
First and second derivative codes of the MITgcm are generated by means of AD. 
The first derivative propagates the uncertainty of controls forward to the target state by a bilinear form product
\begin{equation} \label{eqFWDproj}
\Delta z_{}^2  = {\bf{g}}^T {\bf{P_{xx}g}} 
\end{equation}
of model gradients $\bf{g}$ with control covariance matrix.  
A Lanczos algorithm is applied to extract the leading eigenvectors and eigenvalues of the Hessian matrix
\begin{equation} \label{eqHeig}
 {\bf{H}} = \sum\limits_i {\lambda _i {\bf{v}}_i {\bf{v}}_i^T }
\qquad
\end{equation}
 representing the constrained uncertainty patterns and the inverse of the corresponding uncertainties. 
Since the ocean state estimation problem is a formally ill-posed problem, that is the assimilated observation data is not sufficient to fully constrain all the degrees of freedom of the unknown ocean state, the Hessian is singular and its inverse is expressed in a spectral Moore-Penrose pseudoinverse form
\begin{equation} \label{eqHpinv}
 {\bf{H}}^{+} = \sum\limits_i {\frac{1}{{\lambda _i }}{\bf{v}}_i {\bf{v}}_i^T }.
\end{equation}

This 
allows an explicit analysis of the structure of the Hessian inverse and enables a lossless compression of its computation.  
The dimensionality of the large matrix inversion is reduced and numerical artifacts minimized by explicitly determining  the data-supported subspace of uncertainty, omitting the estimation nullspace unconstrained by the observations. This preserves the resolved uncertainty patterns from numerical distortion by avoiding the regularization procedure, utilized 
to suppress the nullspace in previous implementations of ill-conditioned Hessian inversion \citep{LoschWunsch2003}.

\definecolor{dgreen}{rgb}{0,0.7,0}
\newif\ifcoloreq 
\coloreqtrue     

Two versions of complete inverse-predictive UQ algorithms were developed. One, following the statistical frequentist approach, assimilates observation uncertainty directly, inverting the pure undistorted Hessian and projects the resulting uncertainty of controls to the QoI domain: 
\ifcoloreq 
	\begin{equation} \label{eq_scheme_pure} 
	\begin{array}{lclcl}
	 \textcolor{red}{\bf{R}} &\longrightarrow & \;\color{dgreen}{\bf{P}}_{{\bf{xx}}}=\left({\bf{H}} \right)^{-1} \qquad &\longrightarrow & \textcolor{dgreen}{\Delta z_{}^2}  = {\bf{g}}^T \textcolor{dgreen}{{\bf{P}}_{{\bf{xx}}}} {\bf{g}}
	 \end{array}
	\end{equation}
\else
	\begin{equation} \label{eq_scheme_pure} 
	\begin{array}{lclcl}
	 {\bf{R}} &\longrightarrow & \;\bf{P}_{{\bf{xx}}}=\left({\bf{H}} \right)^{-1} \qquad &\longrightarrow & \Delta z_{}^2  = {\bf{g}}^T {\bf{P}}_{{\bf{xx}}} {\bf{g}}
	 \end{array}
	\end{equation}
\fi 
The other, set in the Bayesian framework, computes the posterior reduction of prior uncertainty and is numerically equivalent to regularized regression regularization technique: 
\ifcoloreq 
	\begin{equation} \label{eq_scheme_prior} 
	\begin{array}{lclcl}
	 ~ & & \color{blue}{\bf{P}}_{\rm{0}} &\longrightarrow & \textcolor{blue}{\Delta z_{}^2}  = {\bf{g}}^T \textcolor{blue}{{\bf{P}}_{\rm{0}}} {\bf{g}} \\ 
	  ~ & &\downarrow  \\ 
	 \textcolor{red}{\bf{R}}  &\longrightarrow &{\bf{P}} = \left( {\textcolor{dgreen}{\bf{H}} + \textcolor{blue}{{\bf{P}}_0^{ - 1}} } \right)^{ - 1}  &\longrightarrow &\Delta z_{}^2  = {\bf{g}}^T {\bf{Pg}} \\ 
	 \end{array}
	\end{equation}
\else
	\begin{equation} \label{eq_scheme_prior} 
	\begin{array}{lclcl}
	 ~ & & \bf{P}_{\rm{0}} &\longrightarrow & \Delta z_{}^2  = {\bf{g}}^T {\bf{P}}_{\rm{0}} {\bf{g}} \\ 
	  ~ & &\downarrow  \\ 
	 \bf{R}  &\longrightarrow &{\bf{P}} = \left( {\bf{H} + {\bf{P}}_0^{ - 1} } \right)^{ - 1}  &\longrightarrow &\Delta z_{}^2  = {\bf{g}}^T {\bf{Pg}} \\ 
	 \end{array}
	\end{equation}
\fi 
The posterior covariance  of controls $\bf{P}$ is computed by inversion of the sum of the misfit Hessian $\bf{H}$ and the inverse of the prior covariance of controls $\bf{P}_{\rm{0}}$. Both the prior and the posterior control uncertainties are projected onto the QoI domain by \eqref{eqFWDproj} resulting in prior and posterior estimates of the Drake Passage transport.  

\begin{figure}[h]
 \centerline{
  \includegraphics[width=16pc]{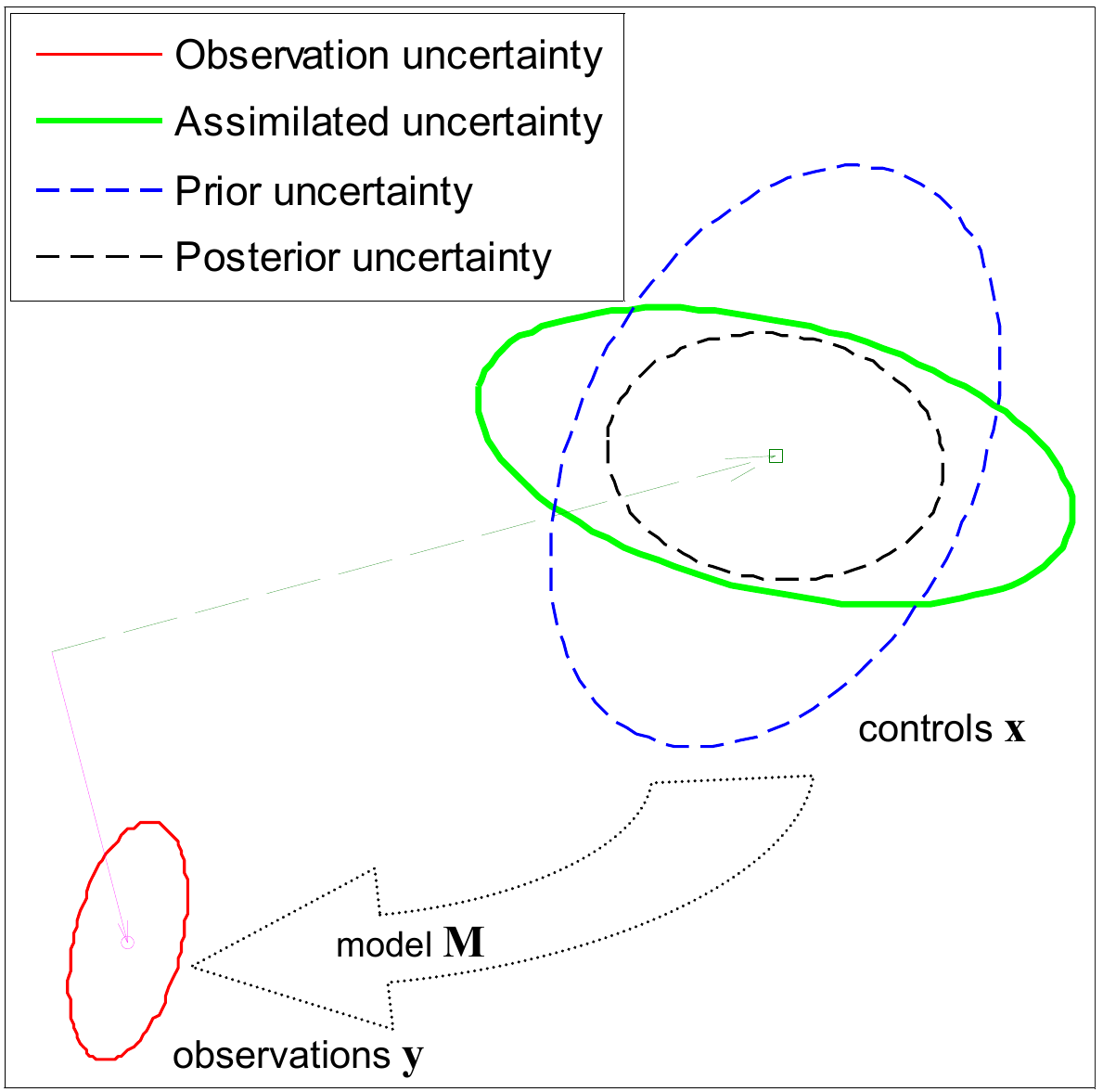}
 }
 \caption{Schematic visualization of inverse uncertainty propagation in two dimensions for a linear invertible forward model \textbf{M}. Pure (green) assimilated uncertainty is compared to prior (blue) and posterior (black) uncertainties of the controls.}
 \label{fig_UQscheme}
\end{figure}



A schematic of both uncertainty assimilation procedures is illustrated in Figure~\ref{fig_UQscheme}. 
The pure Hessian inversion technique \eqref{eq_scheme_pure} inverts the forward model $\textbf{\textit{M}}(\bf{x})$ to obtain the assimilated uncertainty. It is combined with the prior uncertainty in the Bayesian technique \eqref{eq_scheme_prior} to obtain the posterior uncertainty. Implications of the estimation nullspace unconstrained due to the singularity of the Hessian inversion are further discussed in this context in \citet{Kalmikov2014}. 



\section{Mechanisms of uncertainty propagation}

\subsection{Forward uncertainty propagation}
  \label{ssec:FWD_UQ}

\begin{figure*}[t]
 \centerline{
  \includegraphics[width=39pc]
  {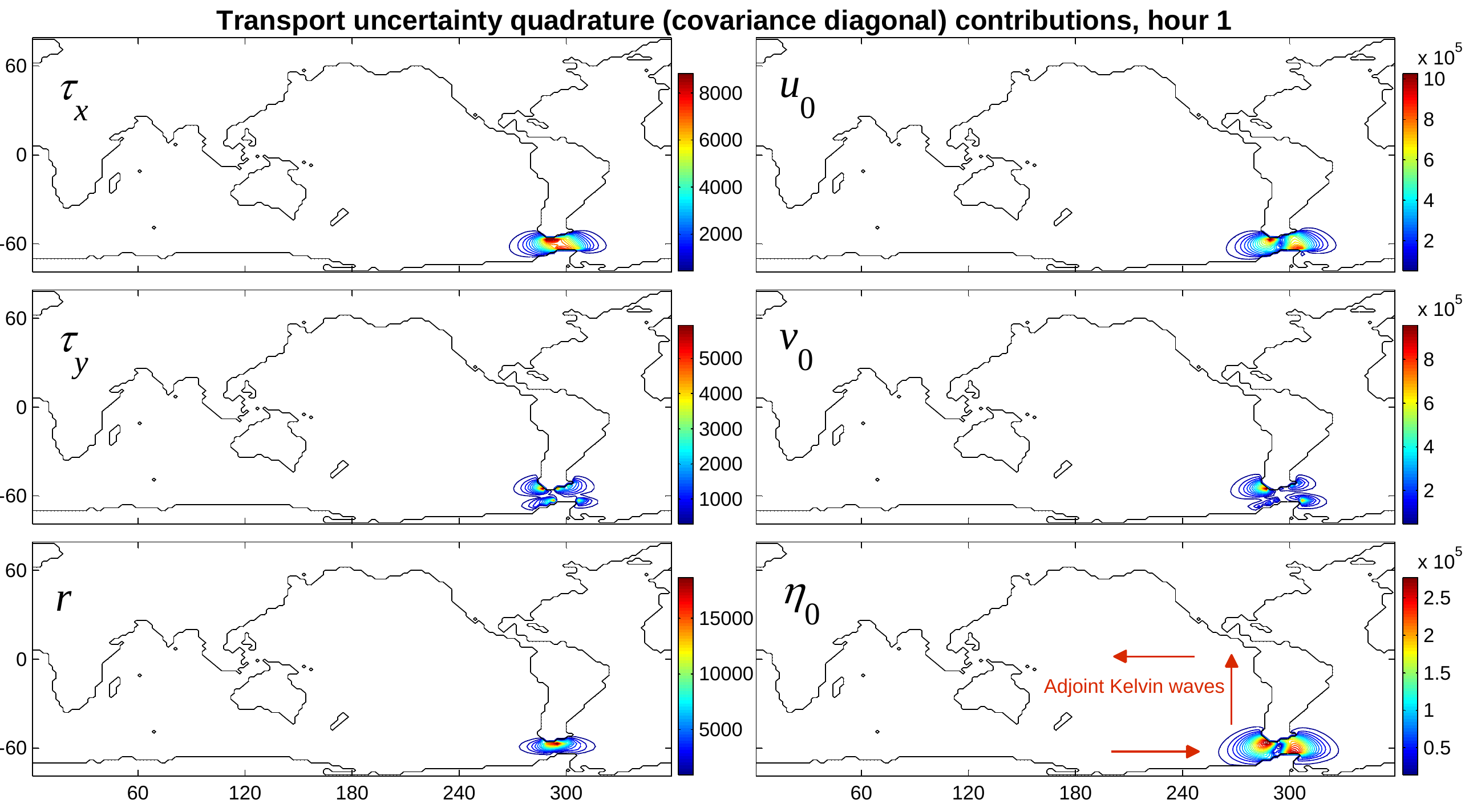}
 }
 \caption{Quadrature contributions (in m$^3$~s$^{-1}$) to uncertainty of Drake Passage transport in forward propagation of the prior uncertainty of controls. The 6 panels correspond to the initial and boundary condition control fields arranged according to Table 1. Results are shown for different forward uncertainty propagation times: 1 hour, 9 hours, 10 days. Sources of Drake Passage transport uncertainty include local effects of zonal wind forcing and bottom friction, upstream meridional wind forcing, and teleconnected effects of initial conditions carried by the adjoint Kelvin waves.}
  \label{fig_QuadratureContributions_1h}
\end{figure*}

\begin{figure*}[t]
 \centerline{
  \includegraphics[width=39pc]{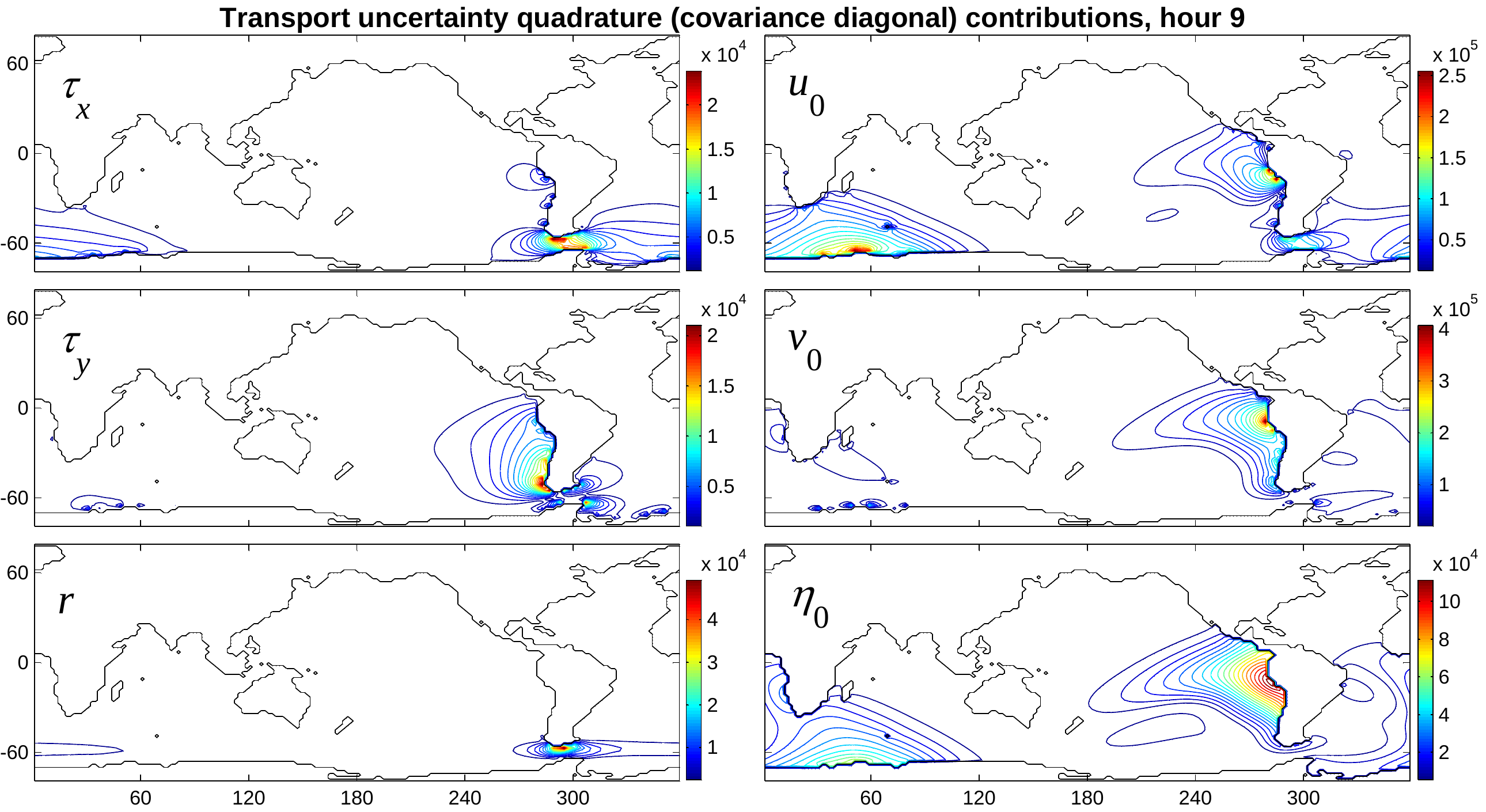}
 }
 \caption{Same as in Figure \ref{fig_QuadratureContributions_1h} but for 9 hours forward uncertainty propagation. Kelvin wave signals extend to the ocean from the American and Antarctic continents in all the initial conditions fields.}
  \label{fig_QuadratureContributions_9h}
\end{figure*}

\begin{figure*}[t]
 \centerline{
  \includegraphics[width=39pc]{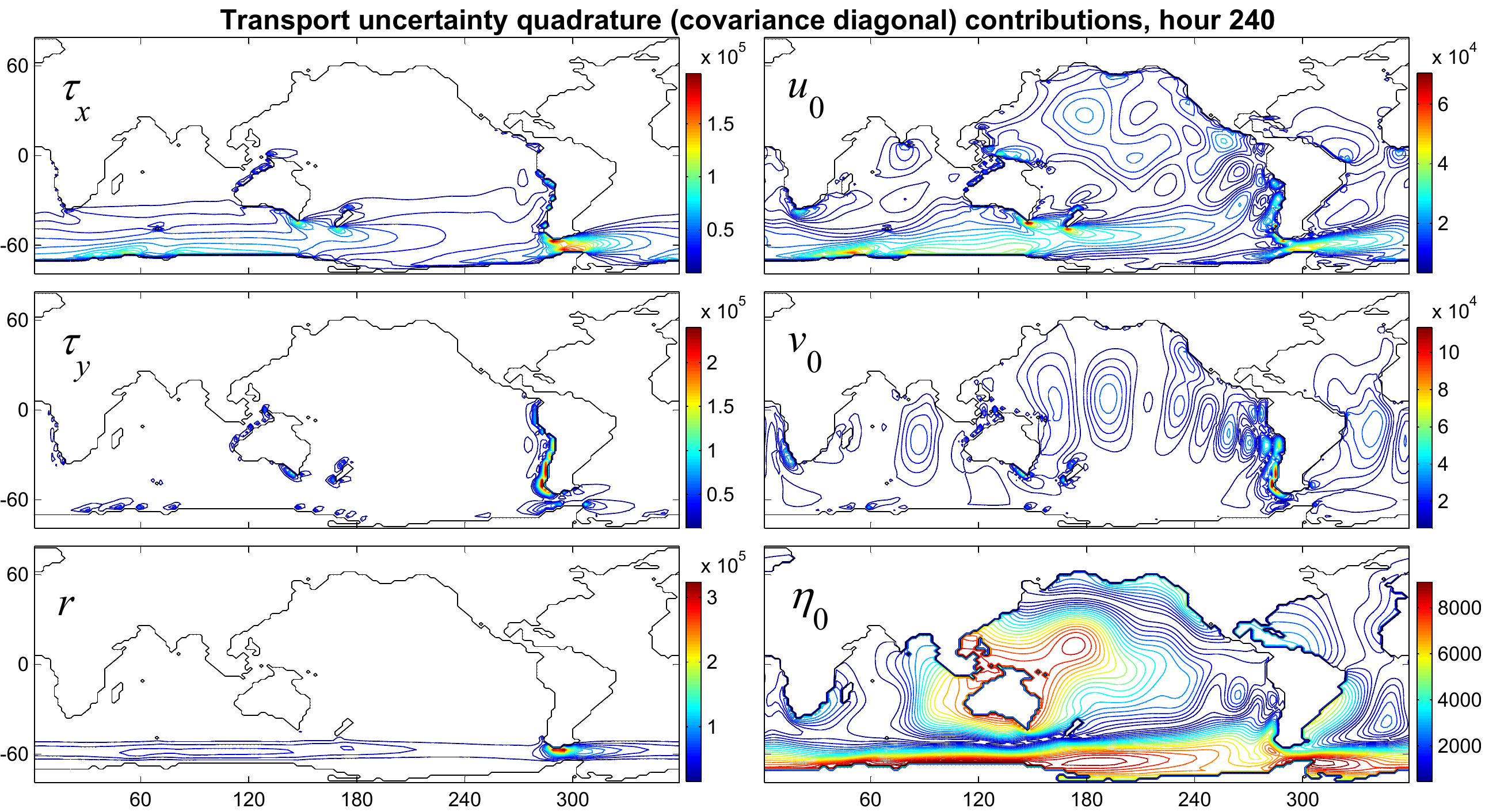}
 }
 \caption{Same as in Figure \ref{fig_QuadratureContributions_1h} but for 10 days forward uncertainty propagation. Rossby waves propagate across the equator from west to east in the initial meridional velocity field (middle right panel).}
  \label{fig_QuadratureContributions_10d}
\end{figure*}

Details of uncertainty dynamics in global ocean modeling are complex and visualization is limited due to the high dimensionality of the multivariate spatio-temporal uncertainty state space\footnote{Note, that uncertainty dimensionality is quadratic in the total number of the physical state space dimensions, required to quantify the cross-coupling of uncertainties for each pair of the discretized model variables. Size of the Hessian matrix is square of the vectorized solution size.}.
Therefore, to facilitate understanding, it is instructive to begin the analysis of uncertainty dynamics with its forward propagation, eqn.~\eqref{eqFWDproj}, which is a deterministic projection of uncertainty from the model control space to the target state, here a single scalar QoI. 
Properties of the control space uncertainties can be examined through the lens of various such projections, by comparing individual control variables` projection components (see below), or by marginalization of control space uncertainties by projection with Cartesian unit vectors.  
Here we analyze the uncertainty of the steady state transport $z \pm \Delta z$ in simulations converged to and reinitialized with steady state initial conditions at time $t_0$. Although simulated transport $z(t)$ is in steady state at all times $t > t_0$, perturbations about this equilibrium are not steady and result in evolving uncertainties $\Delta z(t)$. Our time-resolving UQ method captures this evolution of uncertainty and explains the dynamics of its physical mechanisms, as shown below. 

According to eqn.~\eqref{eq_scheme_prior} either the prior or the posterior uncertainty of controls can be propagated forward. In this section we discuss the simpler case of propagation of the prior uncertainty, assumed uncorrelated between different control vector components and homogeneous for each of the six different fields of the control vector. In other words, the prior covariance matrix is diagonal and the diagonal is given by the squares of the standard errors
 [ $\Delta \tau ^x$, $\Delta \tau ^y$, $\Delta r$, $\Delta u_{t_0}$, $\Delta v_{t_0}$, $\Delta \eta_{t_0}$] =[0.1~Pa, 0.1~Pa, $5.0\times10^{-3}$~m~s$^{-1}$, 0.01~m~s$^{-1}$, 0.01~m~s$^{-1}$, 0.1~m] corresponding to the fields in Table 1.                                                                                         
The general case of non-uniform diagonals and propagation of coupled uncertainties is discussed in section 4d below for the posterior covariance matrix, which is more complicated due to the additional dependence on the assimilation time $t_a$. 

The covariance of controls is forward projected with transport gradient \eqref{eqFWDproj} calculated with the adjoint MITgcm code. For the diagonal prior covariance the scalar form \eqref{eq_ForwardUQ.scalarForm}	of the forward projection reduces to \textit{a quadrature sum} of uncorrelated control uncertainty components 
\begin{equation} \label{eq_quadrature_sum} 
\Delta z_{}^2  = \sum\limits_i {\left( {\frac{{\partial z}}{{\partial x_i }}} \right)} ^2 \Delta x_i^2 
\end{equation}
This summation is the basis of root-sum-square addition rule for independent uncertainties, as applied to Drake Passage transport estimation error analysis by \citet{Cunningham2003}. The sum can be understood as a sensitivity-weighted addition of control variances: the stronger the sensitivity to certain controls -- the larger their contribution to the total transport uncertainty. Note that UQ formalism accounts only for the absolute magnitude of the sensitivity, its sign is ignored, in contrast to the closely related "normalized sensitivity" analysis in \citet{Heimbach2011}. 
%

A simple example is the initial uncertainty of the transport at the reinitialization time $t_0$. 
 There are 5 grid cells spanning the Drake Passage at the longitude of the transport calculation. The cross sectional area 
 of each westerly
 face is $a=1.1\times 10^9$m$^2$. Since the transport is just a sum of the zonal volume fluxes $z= \sum_{i} a_iu_i$, and if the uncertainties of the initial zonal flow are independent, then the uncertainty of the transport is given by addition in quadrature of each of the grid cell contributions
 $\Delta z= a \sqrt{ \sum_{i} \Delta u_i^2 }$. For the initial prior zonal flow uncertainties $\Delta u_{t_0}=0.01$~m~s$^{-1}$ 
the magnitude of the initial prior uncertainty of zonal transport is $\Delta z(t_0)=24.6$~Sv. 
Note, that this simple calculation is possible for the initial time $t_0$ only, because of conditional independence of the transport on the other controls at this time. For other times, uncertainties of other control variables contribute to the uncertainty of the transport as discussed below.



Plotting quadrature sum components for each scalar element of the control vector visualizes the separate factors contributing to transport uncertainty if uncertainties of controls are independent or to single out the contributing factors without considering their correlations. (The limits of this approach are discussed in Section 4d). These contributions are shown in Figures \ref{fig_QuadratureContributions_1h}, \ref{fig_QuadratureContributions_9h} and \ref{fig_QuadratureContributions_10d} 
for different times\footnote{For more times see animations in the supplemental materials --- 24 hour and 10 day uncertainty evolution patterns. }
of forward uncertainty propagation 
$t - t_0$
and are seen to constitute dynamically meaningful patterns evolving in space and time.
Analysis of these patterns and their dynamics allows identification and understanding of the physical mechanisms of forward uncertainty propagation. Since in this case control uncertainties are homogenous in space, the evolution of their contribution patterns is fully explained by evolution of the corresponding sensitivity fields. 

The mechanisms of sensitivity evolution are understood in terms of adjoint time dynamics imposed by time-dependent Jacobian operators calculated with the adjoint model  \citep{Heimbach2011}. 
The adjoint time is the backward lead time $t_0 - t$ from the analysed QoI $z(t)$ to the time of the initial conditions specification $t_0$, i.e. the negative of forward uncertainty propagation time. The adjoint time dynamics describes the evolution of the sensitivity fields such as 
	\begin{equation} \label{eq_sensitivity} 
		\frac{{\partial z(t)}}{{\partial u(t_0) }}
	\end{equation}
for fixed QoI time $t$, tracing backward to $t_0$ the source of QoI perturbations.  
The adjoint exposes several mechanisms at play on different time scales. 
Initially all effects are local -- only the uncertainties at the geographic location of transport calculation contribute (Figure~\ref{fig_QuadratureContributions_1h}). As time proceeds, distinct barotropic signals radiate from the Drake Passage region, propagating through the initial conditions domains in the opposite, time-reversed direction of the corresponding physical waves mechanisms and therefore reflecting the "adjoint dynamics". 
These uncertainty waves propagate backward carrying 
the information on the sources of transport uncertainty of the forward evolving circumpolar circulation. The displayed patterns correspond to adjoint Kelvin waves propagating in the opposite direction to actual Kelvin waves along the Antarctic and South American coasts (Figure \ref{fig_QuadratureContributions_9h}) and crossing the Pacific from east to west.  After about 10 days, adjoint Rossby waves are detected (Figure \ref{fig_QuadratureContributions_10d}) propagating the uncertainty signals across the equatorial Pacific from west to east in the opposite direction to equatorial Rossby waves. 

\begin{figure}[h]
 \centerline{
  \includegraphics[width=19pc]{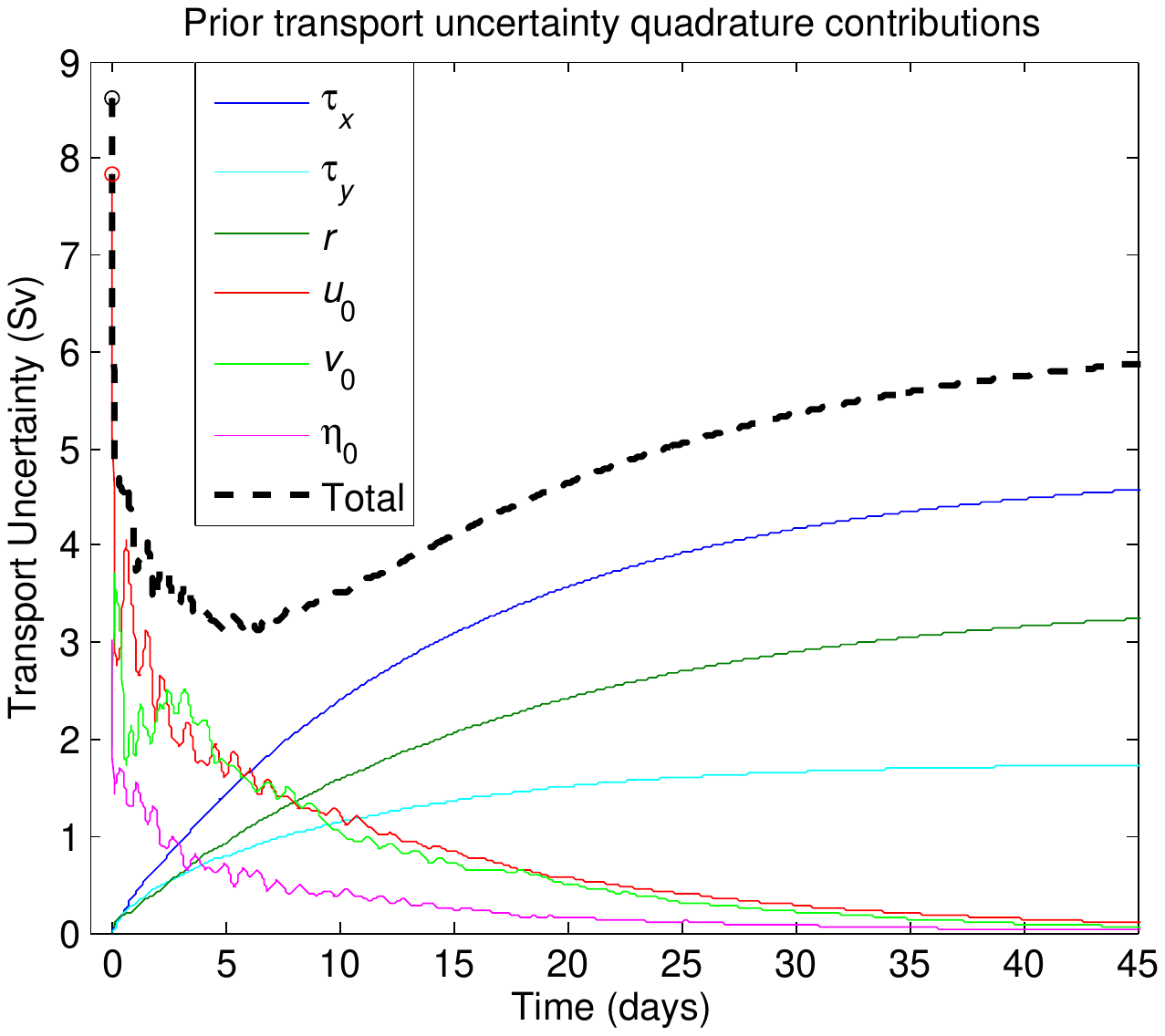}
 }
 \caption{Time evolution of spatially aggregated uncertainty contributions due to each initial and boundary conditions field. Contributions of initial velocity components and surface elevation decay to zero; contribution of boundary conditions -- wind stress components and bottom friction coefficient grow to steady state value. The resultant transport uncertainty evolution is shown with black dashed line, exhibiting superposition of transient decay and asymptotic growth to steady state. Fluctuations are seen to modulate the curve during the initial 10 days, representing oscillating effects of Kelvin and Rossby waves.}
   \label{fig_ForwardEvolution}
\end{figure}

In addition to identifying separate patterns of uncertainty propagation, it is instructive to examine the magnitude of uncertainty contributions from different physical sources. Noting the changes in range represented by the colorbars on the panels of Figures \ref{fig_QuadratureContributions_1h} to \ref{fig_QuadratureContributions_10d} for each two-dimensional field, shows that sensitivity of transport to initial conditions decays to zero in the steady state, while sensitivity due to steady, time-mean boundary conditions grows from zero to steady state values. Initially, close to model reinitialization, the uncertainty of transport is entirely due to the initial condition uncertainty, while at the steady state it is entirely due to the boundary conditions. The major grid-point-level factor in steady state is the uncertainty due to bottom friction, especially on the northern side of Drake Passage (Figure \ref{fig_QuadratureContributions_10d}). A band of bottom friction uncertainty grows from the Passage area to encircle the Southern Ocean in  steady state. Another uncertainty source in steady state is the magnitude of wind forcing, reaching maximum contribution in the Passage area and downstream for the zonal component and concentrating upwind on the western coast of South American continent for the meridional wind stress. While the local sources of uncertainty at the Passage are intuitive, the nonlocal boundary condition effects provide useful insights, such as that downstream zonal wind stress is more important for constraining the Drake Passage transport than upstream. 

Summing all quadrature uncertainty contributions \eqref{eq_quadrature_sum} results in the prior estimate of uncertainty of Drake Passage transport. For the lead time 9 hours shown in Figure~\ref{fig_QuadratureContributions_9h} the standard error is $\Delta z=4.6$~Sv. The magnitude of the simulated zonal transport is 100~Sv, therefore its uncertainty is $\pm 4.6$\%. Repeated for different times of forward uncertainty propagation 
the result traces the forward time evolution of transport uncertainty, shown in Figure \ref{fig_ForwardEvolution} together with partial quadrature sums for each of the 6 control fields integrated across each two-dimensional map. The total uncertainty (thick dashed line) is shown to evolve significantly: decreasing precipitously during the initial action of the fast adjoint gravity waves and growing steadily to a steady state after $O(50~\mathrm{days})$. 
The specific shape of the evolution curve -- the depth of the transient minimum and the level at the steady state, depends in the assumed prior uncertainties of the controls \citep[p. 123]{Kalmikov2013}. 
This is an important result showing that even for simple fixed uncertainties of model initial and boundary conditions and for a simulation of a steady state flow, the uncertainty of the solution is unsteady and exhibits non\-trivial dynamics. 
The uncertainty evolution can be fully explained as superposition of opposite effects of initial and boundrary condition uncertainties. The initial transport uncertainty drop is due to the exponential decay of initial conditions (velocity components and sea surface height) uncertainty effects on transport. In contrast, the steady state transport uncertainty asymptote is entirely due to the (steady, time-mean) boundary condition (wind forcing and bottom friction) uncertainty contributions growing and saturating to steady state values.

\subsection{Inverse uncertainty propagation}
   \label{ssec:INV_UQ}

Similar time-dependence is reflected also in inverse uncertainty propagation, although in a more intricate way. Inverse uncertainty propagation projects uncertainty of the assimilated observations onto model controls by inverting the Hessian of model-observation misfit \eqref{eq_scheme_pure}. Here, the assimilated observations are   surface height data $\eta(t_a)$ from satellite altimetry over a rectangular area at the Drake Passage (Figure \ref{fig_configurationDomain}), specified with independent and homogeneous uncertainties with standard error 
 $\Delta \eta =0.01~\rm m$.
The observations are synthetic, generated in identical twins simulations with the same model, and are available at different forward model times. While the values of these observations are constant, since the forward model is in a steady state, they project varying information on model controls when assimilated at different times $t_a$. In realistic applications, assimilation of data and its uncertainty constrains the model by a superposition of different assimilation times` effects. Here, however, we consider these times separately and analyze uncertainty assimilation as function of inverse propagation time $t_0 - t_a$. The goal is to assess how the dynamical processes resolved by the model transform the assimilated altimetry information to constrain the integrated volume transport. Tracing the evolution of the Hessian structure for different assimilation times reveals the barotropic mechanisms of inverse propagation of the assimilated uncertainty. 
 
 \begin{figure}[b]
 \centerline{
  \includegraphics[width=19pc]{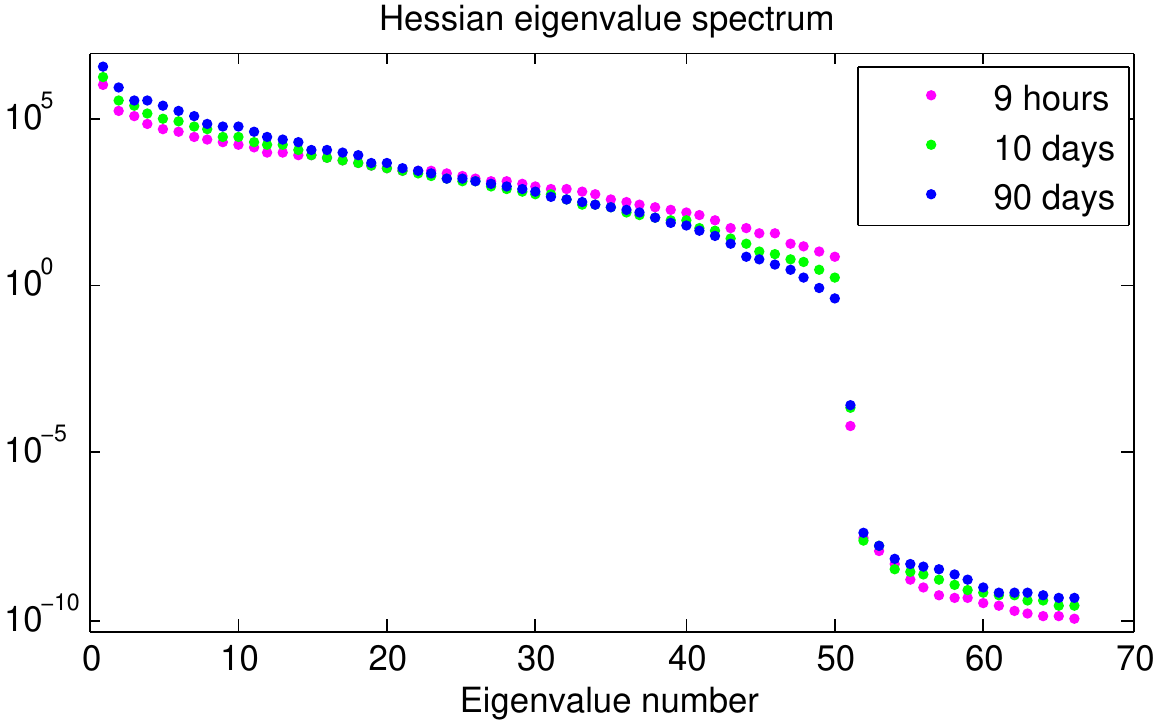}
 }
 \caption{Hessian eigenvalue spectra for assimilated sea surface observations at the Drake Passage, shown for 9 hours, 10 days and 90 days inverse uncertainty propagation simulations.}
  \label{fig_eigenvalueSpectra}
\end{figure}


 Analyzing physical processes in a numerical ocean model while minimizing numerical artifacts and resolving a dynamically consistent signal is a priority in ECCO ocean state estimation method \citep{WunschHeimbach2013}. The prior-independent spectral Hessian inversion scheme \eqref{eq_scheme_pure} 
 achieves this by resolving pure \citep[see][]{Kalmikov2014} 
 inverse uncertainty propagation signals, that is by avoiding regularization distortion and bias due to the prior in the Bayessian scheme \eqref{eq_scheme_prior}, 
 and maintaining consistency with forward uncertainty propagation \citep[eq. 4.1.17]{Kalmikov2013}. 
 %
For each observation time the assimilated information is propagated to constrain the data-supported sub-domain of model controls, forming a finite confidence region ellipsoid in the \textit{range space} of the estimation problem. The remaining unconstrained sub-domain is the \textit{nullspace} of the inverse problem, where assimilated observations leave the uncertainty unbounded and no error covariance matrix exists. The spectral Hessian UQ method \eqref{eqHpinv} identifies the structure of the constrained controls, constructs an error covariance reduced to the constrained sub-domain and expresses the confidence ellipsoid in terms of its principal axes given by the eigenvectors of the Hessian. These eigenvectors represent the principal uncertainty patterns spanning the constrained uncertainty assimilation sub-space. The standard errors of these patterns and the lengths of the ellipsoid axes are given by the inverse square roots of the corresponding Hessian eigenvalues. 
 

 The eigenvalues of the Hessian represent the information gained by assimilating uncertain observations. The eigenvalue spectrum (Figure \ref{fig_eigenvalueSpectra}) exhibits a sharp cut-off after the leading 50 eigenvalues. This means that the information gained is concentrated in few orthogonal modes of uncertainty, while the rest of the spectrum constitutes the unconstrained nullspace. 
 Note, this clear separation of the data supported sub-space from the nullspace is enabled by the synthetic observations in the zero-residial configuration of the identical-twins UQ experiment. In realistic state estimation configurations \citep{Kalmikov2014}, the residual misfit contributes a nonlinear Hessian term \eqref{Hessian_eq} 
 to the Hessian matrix which masks the low-rank structure of the so-called "linearized Hessian" term \citep{LoschWunsch2003}, also known as the Gauss--Newton term.
 \begin{equation} \label{Hessian_eq} 
{\bf{H}} = \left( {\frac{{\partial \textbf{\textit{M}}}}{{\partial {\bf{x}}^T }}} \right)^T {\bf{R}}^{ - 1} \left( {\frac{{\partial \textbf{\textit{M}}}}{{\partial {\bf{x}}^T }}} \right) + \left( {\frac{{\partial ^2 \textbf{\textit{M}}}}{{\partial {\bf{x}}\,\partial {\bf{x}}^T }}} \right)^T {\bf{R}}^{ - 1} \left( {\textbf{\textit{M}}({\bf{\hat x}}) - {\bf{y}}} \right)
\end{equation} 
 Eigenvalue spectra for observation uncertainty assimilated 9 hours, 10 days and 90 days after steady state initialization are compared in Figure \ref{fig_eigenvalueSpectra}. The spectra are not stationary and appear to steepen as function of uncertainty assimilation time: for 90 days the leading eigenvalues are larger and the trailing eigenvalues are smaller than for shorter assimilation times. This uncertainty spectrum evolution is a manifestation of redistribution of assimilated information along the scales, suggesting that the information gain for longer inverse uncertainty propagation is more concentrated at larger spatial scales, while for shorter inverse uncertainty propagation -- information gain is comparatively more uniform across the spectrum.

\begin{figure*}[t] 
 \centerline{\includegraphics[width=39pc]
 {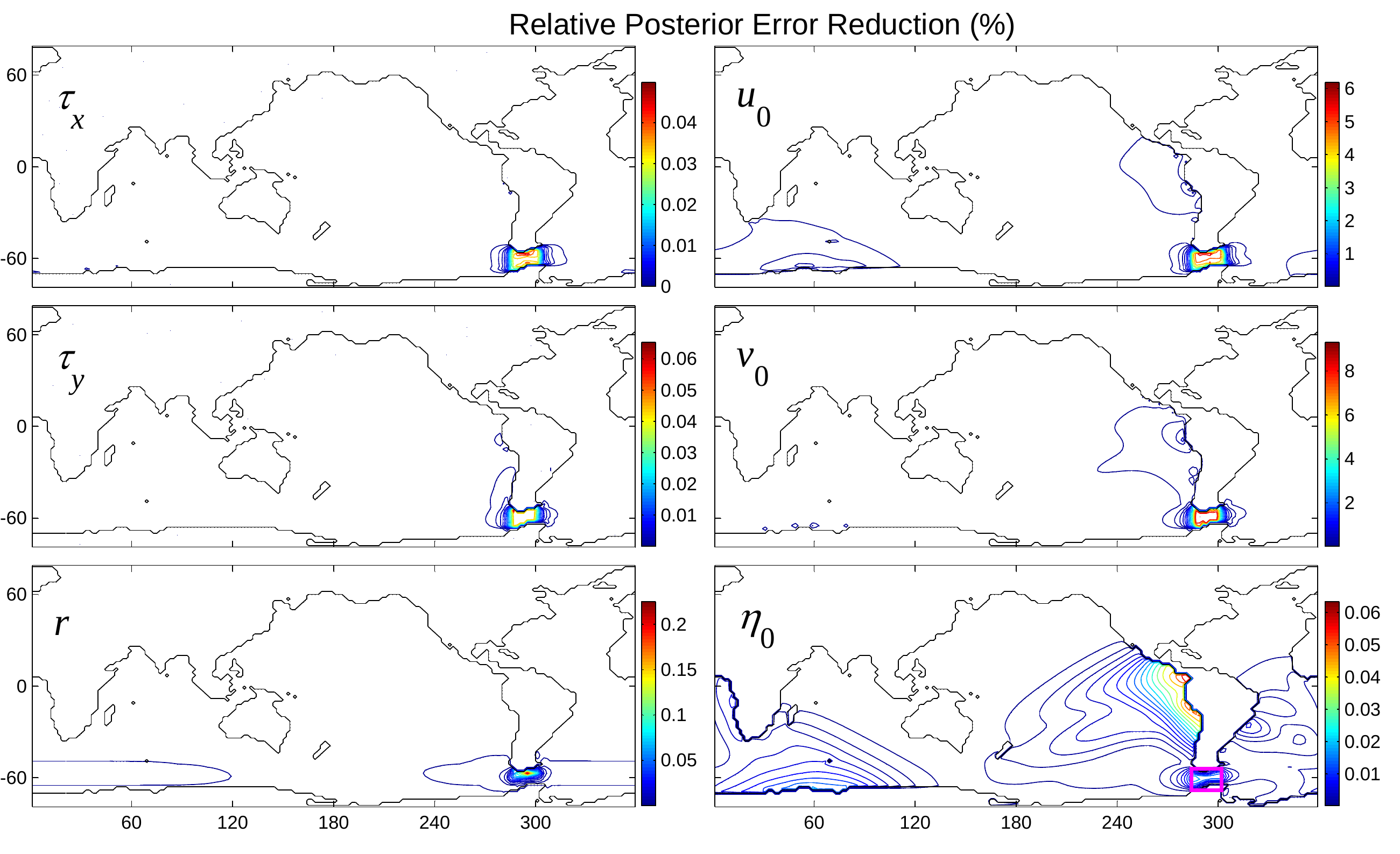}}
  \caption{
  Relative posterior uncertainty reduction (shown in \%) of the 6 control fields due to assimilation of sea surface observations at the Drake Passage (shown with magenta rectangle on right bottom panel) for 9 hour inverse uncertainty propagation simulation. Control vector fields configuration is according to Table 1. Only the diagonal covariance matrix terms can be shown in a single 6 panel figure, because display of the complete posterior uncertainty covariance would require $O(10^5)$ figures like this.} 
  \label{fig_6panel_HessianUQreduction-9hour}
\end{figure*}

\begin{figure*}[t]
 \centerline{\includegraphics[width=39pc]
 {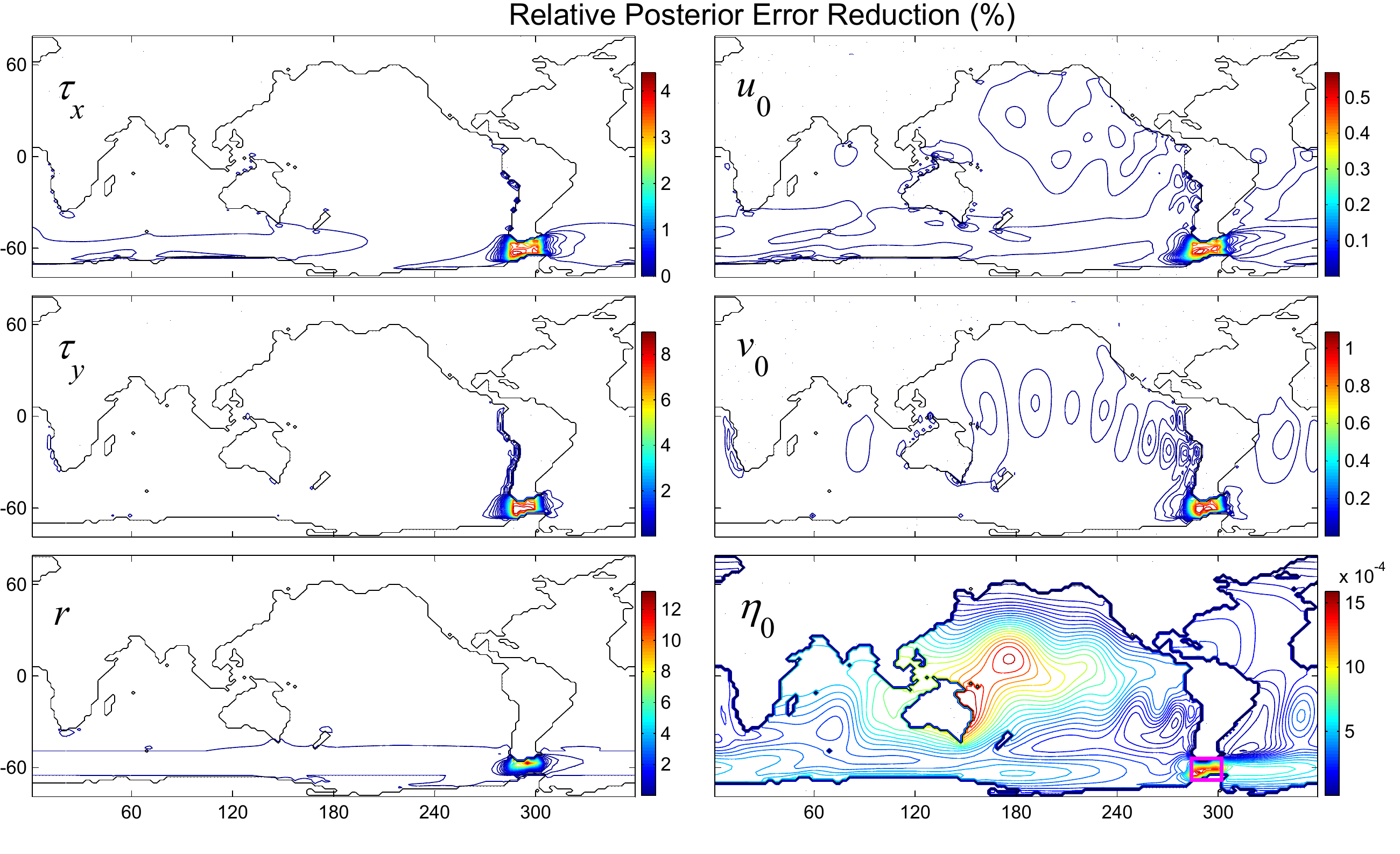}}
  \caption{Same as Figure \ref{fig_6panel_HessianUQreduction-9hour} but for 10 day inverse uncertainty propagation simulation.} 
  \label{fig_6panel_HessianUQreduction-10day}
\end{figure*}

\subsection{Posterior uncertainty reduction of model controls}
   \label{ssec:Post_controls}

Inverse uncertainty propagation is more straightforward to analyze in the Bayesian framework \eqref{eq_scheme_prior}, where uncertainty constrained by observations in the data-supported sub-domain is fused with prior uncertainty of model controls and results in reduced posterior uncertainty. The main advantage is  the existence of a non-degenerate covariance matrix expressing finite uncertainties over the full dimension of the control space. All components of the control vector are constrained and the nullspace is suppressed, the prior regularizes the inverse problem. The covariance describes the structure of a full rank confidence ellipsoid of model controls. Due to the difficulty of visualizing this structure in $O(10^5)$ dimensions we restrict the analysis to the marginal uncertainties of each control vector component, that is the diagonal of the covariance. The effects of the off-diagonal terms are accounted for in the following section by projecting the covariance onto the scalar target domain of volume transport.

The rate of marginal prior-to-posterior uncertainty reduction of controls
\begin{equation} \label{eqUreductionRate} 
\gamma \, = \, \frac{\Delta_{prior}-\Delta_{post}}{\Delta_{prior}}\cdot100
\end{equation}
is shown in Figures  \ref{fig_6panel_HessianUQreduction-9hour} and \ref{fig_6panel_HessianUQreduction-10day}. The uncertainty is reduced in all control variables, not only in the variable where observations were assimilated. This indicates the dynamical cross-coupling between the different control variables. Moreover, the reduction is not limited to the rectangular area of the assimilated observations, but extends beyond it, indicating the cross-coupling within each separate control field. The patterns and evolution of reduction reflect the identified barotropic uncertainty mechanisms. In particular, note that for the observed field, the surface elevation $\eta$, relatively negligible uncertainty reduction is achieved in the geographical area of data assimilation and most of the effect propagated away carried by the adjoint uncertainty waves radiating from the Drake Passage. For the other control fields, in contrast, most of the uncertainty reduction occurs in the Drake Passage region. Weak reduction is seen for zonal flow carried by the adjoint Kelvin waves along the Antarctic coast and for meridional flow along the South American continent after 9 hours (Figure~\ref{fig_6panel_HessianUQreduction-9hour}) and by the adjoint Rossby waves after 10 days (Figure~\ref{fig_6panel_HessianUQreduction-10day}). Uncertainty reduction of the forcing fields is mostly concentrated in the data assimilation area, although the southward wind stress uncertainty is slightly reduced upstream of the Drake Passage as wind forces the flow around the continent and the bottom drag uncertainty reduction follows the circulation zonally around the globe. After 10 days uncertainty reduction is much more significant in the boundary conditions fields, while in the initial conditions it weakens and decays to zero after about 90 days. Assimilation of data in the steady state informs, and therefore constrains, the boundary conditions that govern the steady state. The initial conditions are forgotten by the model in the steady state. They are constrained only by data from the initial 
adjustment period. During that time the constraint of the boundary conditions is much weaker, as it grows to the steady state.

\subsection{Posterior uncertainty reduction of Drake Passage transport}
   \label{ssec:Post_transport}
   
Projecting the posterior uncertainties of the constrained controls to the target variable domain involves combining inverse and forward uncertainty propagation. This results in double\footnote{
 In general non-autonomous systems the dependence is triple, including also the initialization time $t_0$, but here the system is governed by autonomous PDEs invariant under $t_0$ time shift.} 
time dependence of posterior transport uncertainty 
\begin{equation} \label{eqFWDprojTime}
\Delta z_{}^2(t,t_a)  = {\bf{g}}^T(t) {\bf{P}}(t_a) {\bf{g}}(t) .
\end{equation}
For any given assimilation time $t_a$, the forward propagation of the constrained uncertainty evolves in time $t$. For different assimilation times this results in a family of forward evolving posterior uncertainty curves. Each posterior curve can be compared to the prior uncertainty evolution curve, which is independent of uncertainty assimilation time and was analyzed in Section 4a
. The rate $\gamma$ of prior-to-posterior uncertainty reduction \eqref{eqUreductionRate} is computed for each posterior uncertainty curve. The resulting family of uncertainty reduction curves allows a normalized comparison of uncertainty reduction evolution profiles between different data assimilation times.  

%

\begin{figure*}[t]
 \centerline{\includegraphics[width=39pc]{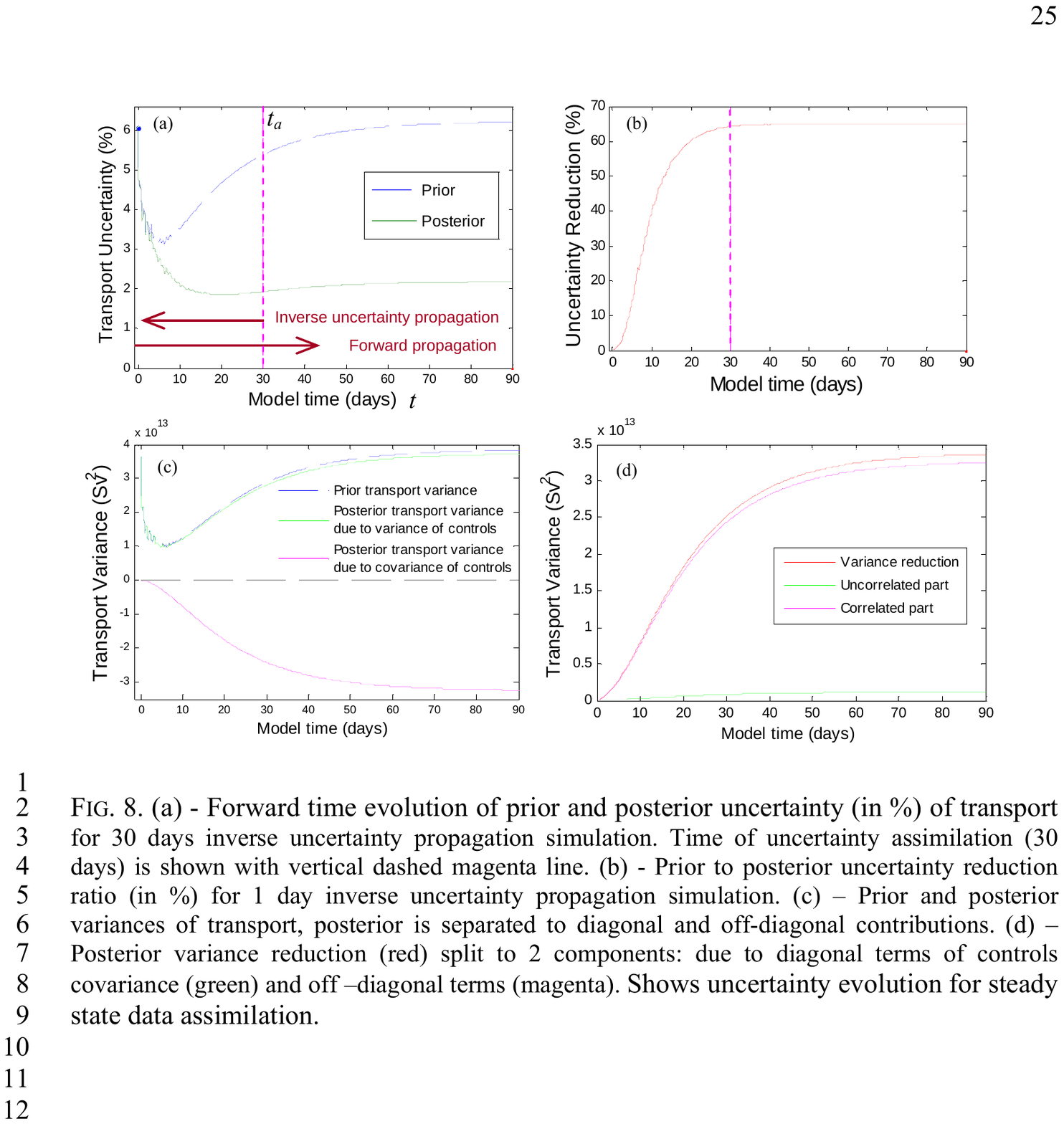}}
  \caption{
 (a) -- Forward time evolution of prior and posterior uncertainty (in \%) of transport for 30 day inverse uncertainty propagation simulation. Time of uncertainty assimilation (30 days) is shown with vertical dashed magenta line. 
 (b) -- Prior to posterior uncertainty reduction ratio (in \%) for 1 day inverse uncertainty propagation simulation. 
 (c) -- Prior and posterior variances of transport, posterior is separated to diagonal and off-diagonal contributions. 
 (d) -- Posterior variance reduction (red) split to 2 components: due to diagonal terms of controls covariance (green) and off–diagonal terms (magenta). Shows uncertainty evolution for the steady state data assimilation.
} 
  \label{fig4panels_assimilate30days}
\end{figure*}

The double time dependence on inverse and forward uncertainty propagation is illustrated in Figure~\ref{fig4panels_assimilate30days}a for assimilation time 30 days, shown by a vertical dashed line. The posterior uncertainty curve drops together with the prior in the first 10 days, but after the prior rebounds and approaches a steady state of 6\% the posterior remains low, rebounding minimally, and asymptotes to about 2\% in steady state. The arrows illustrate the propagation of assimilated uncertainty though the model: from assimilation time at 30 days to controls at time zero\footnote{Here all times are shown relative to the initialization time $t_0$.} via inverse propagation, and by forward propagation from time zero to 90 days. 

\begin{figure*}[t]
 \centerline{\includegraphics[width=39pc]{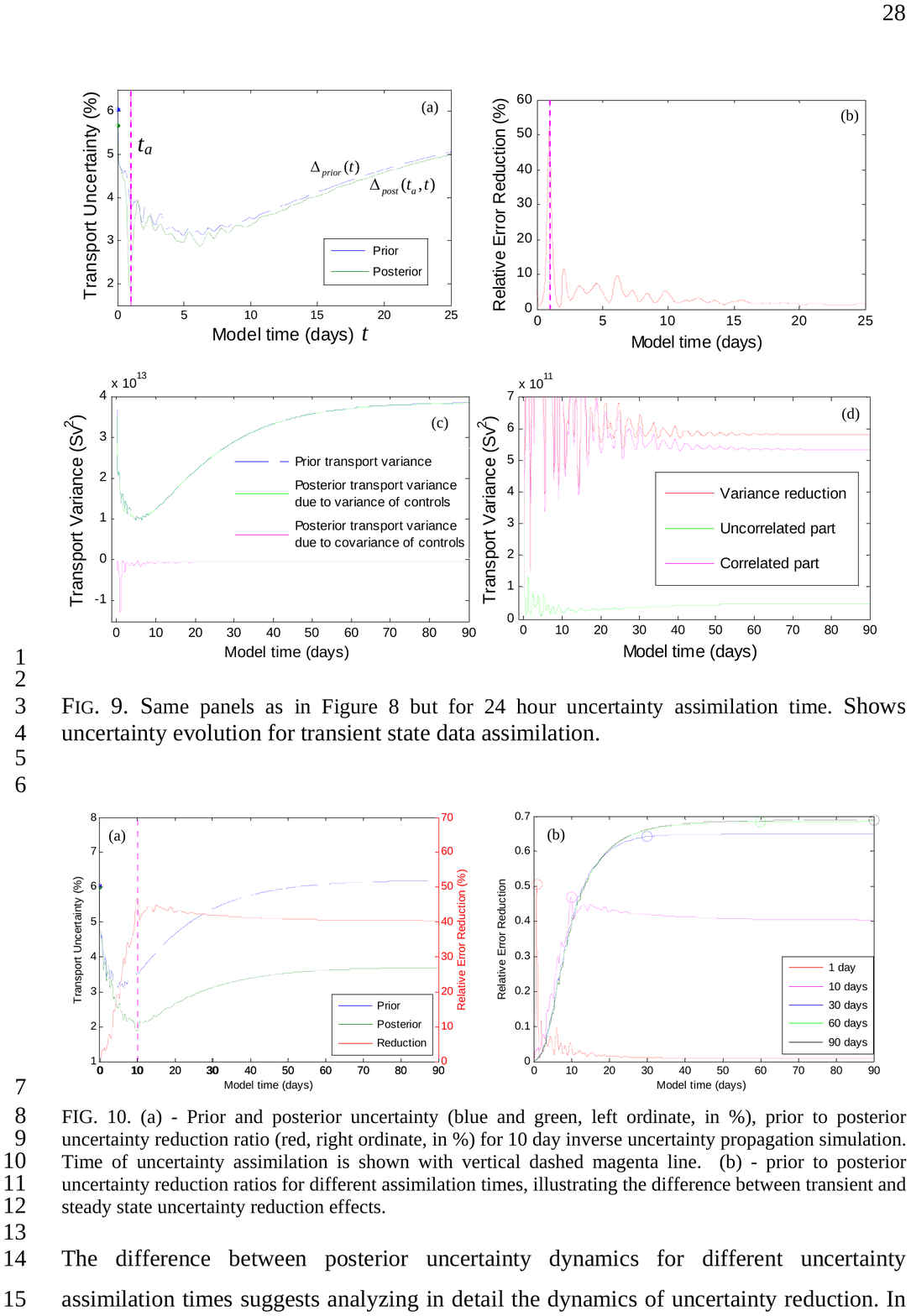}}
  \caption{
  Same panels as in Figure~\ref{fig4panels_assimilate30days} but for 24 hour uncertainty assimilation time. Shows uncertainty evolution for the transient state data assimilation.
} 
  \label{fig4panels_assimilate24hours}
\end{figure*}

The fraction of uncertainty reduction is shown on Figure~\ref{fig4panels_assimilate30days}b -- growing from zero to 65\% in the steady state. This dynamics of uncertainty reduction is very different for other assimilation times. For assimilation time 24 hours (Figure~\ref{fig4panels_assimilate24hours}) the posterior reduction is only transient around assimilation time. The posterior curve follows the prior curve closely for all times (Figure~\ref{fig4panels_assimilate24hours}a), except for dropping precipitously at 24 hours. The rate of uncertainty reduction reaches over 50\% temporarily, but only 1\% reduction remains in the steady state (Figure~\ref{fig4panels_assimilate24hours}b). These different evolution profiles can be explained by the governing barotropic dynamics of the system, which forgets the initial conditions exponentially fast and adjust to a steady state defined by the boundary conditions. When assimilated observations are from the steady state they constrain the boundary conditions, propagating backward for long enough time for the transient effects to subside and steady state to settle. When these contained boundary conditions are propagated forward -- \textit{a stationary uncertainty reduction} results. 
When observations belong to the transient regime, assimilation constraint is limited to initial conditions and the consequent forward propagation reduces uncertainty only temporarily -- a case of \textit{transient uncertainty reduction}. 

\begin{figure*}[t]
 \centerline{\includegraphics[width=39pc]
            {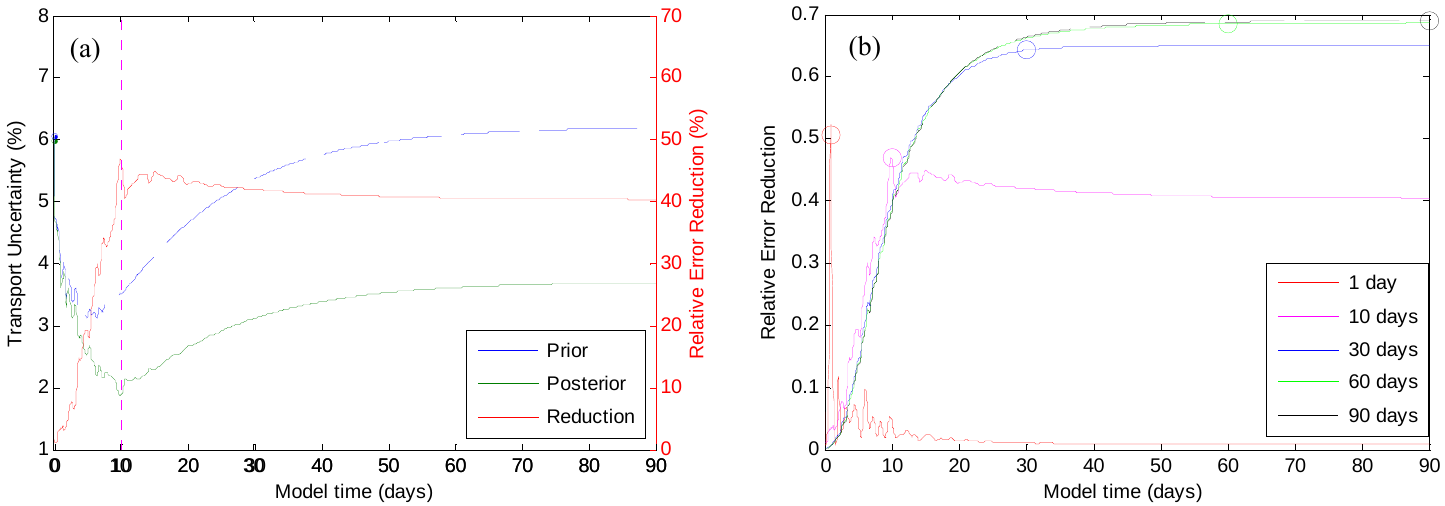}}
  \caption{
  (a) - Prior and posterior uncertainty (blue and green, left ordinate, in \%), prior to posterior uncertainty reduction ratio (red, right ordinate, in \%) for 10 day inverse uncertainty propagation simulation. Time of uncertainty assimilation is shown with vertical dashed magenta line.  
  (b) - prior to posterior uncertainty reduction ratios for different assimilation times, illustrating the difference between transient and steady state uncertainty reduction effects.
} 
  \label{fig2panels_10days-allDays}
\end{figure*}

Superposition of steady and transient uncertainty assimilation effects is shown in Figure~\ref{fig2panels_10days-allDays}a for assimilation time 10 days. A transient uncertainty reduction peak of 47\% occurs at time 10 days, but also permanent uncertainty reduction occurs asymptoting 40\% in the steady state. A family of uncertainty reduction curves is shown in Figure  \ref{fig2panels_10days-allDays}b, with assimilation times marked by circles. The transient peak reduction disappears for assimilation times $t_a$ longer than 10 days.  The steady state reduction grows monotonically with $t_a$ and converge to 70\% for $t_a=90$ days. 
This understanding of uncertainty reduction dynamics has important implications to design of ocean observation sensors. The temporal and spatial spacing of observations should constrain the stationary system regimes or else only a transient uncertainty reduction will be achieved. For highly non-autonomous systems, e.g. forced by broad-spectrum winds, the temporal separation of observations should account for the characteristic uncertainty response times to keep errors continuously low relying only on transient uncertainty reduction.  

%

The difference between posterior uncertainty dynamics for different uncertainty assimilation times suggests analyzing in detail the dynamics of uncertainty reduction. In Section 4a we investigated the separate contributions to the uncertainty of transport given by the quadrature rule \eqref{eq_quadrature_sum}. That mapping from uncertainties of controls to the target variable \eqref{eqFWDproj} was simplified because the prior control uncertainties were not correlated. This is not the case for the posterior uncertainties of controls, because uncertainty assimilation introduces off-diagonal covariance terms to the posterior control covariance matrix. The posterior mapping from controls to the target variable uncertainty is given in scalar form by 

\begin{equation} \label{eq_ForwardUQ.scalarForm}
\Delta z_{}^2  = \sum\limits_i {\left( {\frac{{\partial z}}{{\partial x_i }}} \right)} ^2 \Delta x_i^2  + \sum\limits_i {\sum\limits_{j \ne i} {\frac{{\partial z}}{{\partial x_i }}\frac{{\partial z}}{{\partial x_j }}\Delta x_i \Delta x_j \rho _{ij} } }
\end{equation}

Only the first sum (quadrature) of sensitivity weighted uncertainties was considered in the previous analysis. 
Results were shown in terms of the spatial structures of the separate control fields contributions for each forward propagation time, as well as time series of the evolution of the separate contributions. 
It is much more technically challenging to conduct similarly detailed analyses to the covariance contributions to the target uncertainty. 
The double summation includes cross products of sensitivity fields weighted by the posterior covariances of controls, or equivalently, the cross products of sensitivity-weighted posterior uncertainties scaled also by their correlation coefficients. The number of terms in the double summation is square of the number of quadrature sum terms and therefore cannot be simply visualized for the large dimensionality of the ocean state estimation problem. 

The posterior quadrature 
contribution patterns can be plotted, just as in Section 4a, however they are not shown here because visually they are almost indistinguishable from the prior quadrature uncertainties (e.g., Figure~\ref{fig_QuadratureContributions_9h}). This can be explained by the relatively small reduction of the diagonal terms in the posterior covariance matrix of controls, as is shown quantitatively on Figures~\ref{fig_6panel_HessianUQreduction-9hour} and \ref{fig_6panel_HessianUQreduction-10day}. 
This is confirmed by plotting only the posterior quadrature sum term (the sum of squares, not root of sum of squares) 
as function of forward uncertainty propagation time in panels (c) of Figures \ref{fig4panels_assimilate30days} and \ref{fig4panels_assimilate24hours}. The weighted sum of uncorrelated (marginal) variances is practically indistinguishable (green line) from the prior variance (blue line) for 24 hours assimilation (Figure~\ref{fig4panels_assimilate24hours}c) and only slightly smaller for 30 days (Figure~\ref{fig4panels_assimilate30days}c). 
On the other hand, the other part of the forward projection sum \eqref{eq_ForwardUQ.scalarForm} -- the double weighted sum of the correlated posterior covariance components $\Delta x_i \Delta x_j \rho_{ij}$ is very significant. 
It (magenta line) fully explains both the steady state and the transient uncertainty reductions, growing to 
negative steady state on Figure~\ref{fig4panels_assimilate30days}c and spiking to transient negative peak on Figure~\ref{fig4panels_assimilate24hours}c. 
Note, that while the sum of variances is always a positive number (first sum in eqn.~\ref{eq_ForwardUQ.scalarForm}), the sensitivity-weighted sum of off-diagonal covariances can be negative when either the sensitivities have opposite signs, like sensitivities to wind forcing and bottom friction, or when the correlation coefficient $\rho_{ij}$ is negative, thus leading to an overall uncertainty reduction.
The same conclusion is illustrated in terms of uncertainty variance reduction fractions --- i.e. \eqref{eqUreductionRate} but for variances instead of standrard errors --- on Figure~\ref{fig4panels_assimilate30days}d for 30 days assimilation and Figure~\ref{fig4panels_assimilate24hours}d for 24 hours. 
In both cases the reduction due to the uncorrelated part (green lines) is much smaller than the posterior variance reduction due to correlations (magenta lines). Because the prior covariance of controls is diagonal, all nonzero off-diagonal terms in the posterior covariance matrix are solely due to the uncertainty assimilation. The interpretation is that the assimilation of observation uncertainty and its inverse propagation through the model cross-couples the control fields and this coupling is responsible for the bulk of uncertainty reduction. An explicit example of such coupling is the correlation of wind forcing and bottom friction boundary conditions in the steady state posterior \citep[detailed derivation in Section 4.1][]{Kalmikov2013}. 
 While the correlation is positive, the sensitivity-weighted product 
\begin{equation} \label{eq_FWD.scalarCrossTerm}
 \frac{{\partial z}}{{\partial \tau ^x}} 
  \frac{{\partial z}}{{\partial r}}
  \Delta \tau ^x \Delta r \, \rho _{\tau ^x r} 
\end{equation}
is negative, reducing the posterior uncertainty of the transport.

This analysis highlights the importance of uncertainty correlations between control variables, in contrast to uncertainty summation via quadrature (root-sum-squares) rule, which neglects uncertainty correlations and leads to highly misleading results. We have shown that off-diagonal covariance terms 
dominate posterior uncertainty reduction. Physically, it is the coupling of ocean fields that contains mutual information and allows uncertainty reduction in ocean state estimation. The components of the ocean system are dynamically connected 
and the ability of our uncertainty assimilation method to resolve their uncertainty correlations is key to quantification of the posterior uncertainty reduction of Drake Passage transport.

\section{Summary and Conclusion}

Barotropic mechanisms of uncertainty propagation associated with estimation of Drake Passage transport are analyzed by applying the Hessian uncertainty quantification method to a global barotropic model of the ACC. The formal mathematical procedure projects error covariance matrices of surface observations to uncertain GCM boundary and initial conditions as well as numerical model parameters, and which, together, form a high-dimensional control space. Data-constrained control error covariance information is then projected to the scalar QoI, here Drake Passage transport, quantifying its estimation error separately from its temporal variability. 
This procedure is developed to address the need to formally assess Drake Passage transport estimation uncertainty, and complements analyses of its temporal variability in the SOSE solution \citep{Mazloff_etal2010}. The framework is a prototype of  a scalable UQ machinery to the ECCO ocean state estimation system \citep{Wunsch2007}. 

Targeting the kinematically and dynamically consistent state estimation method of ECCO \citep{WunschHeimbach2013}, the Hessian method for UQ was designed to preserve this dynamical consistency and eliminate the numerical artifacts in the resolved uncertainty dynamics \citep{Kalmikov2014}. 
Two versions of uncertainty propagation algorithms are demonstrated: one evaluates reduction of prior uncertainties, another does not require prior assumptions. Both versions combine inverse and forward uncertainty propagation for assimilating observation uncertainties in the model and for constraining the estimated QoI. 

The analysis of uncertainty propagation in the model is time-resolving, capturing the dynamics of uncertainty evolution and revealing transient and stationary uncertainty regimes. 
The results expose the dynamical links between assimilated altimetry observation uncertainties and the spatio-temporal structure of the uncertainty of the estimated ocean state. We explain inverse and forward uncertainty propagation mechanisms in terms of barotropic response of global ocean to wind forcing in presence of bottom friction. By resolving the time dimension of uncertainty dynamics, the system distinguishes between forward and backward in time uncertainty evolution, explained physically in terms of normal versus adjoint dynamics. Dynamical coupling of uncertainty between different physical fields is explained by the off-diagonal uncertainty covariance structure and evolution. Teleconnections in global ocean fields and cross-coupling between different ocean state variables are identified. 

Uncertainty reduction is demonstrated in both model controls and the circumpolar transport. The effects of controls uncertainty reduction due to decrease of diagonal covariance terms are compared to dynamical coupling of controls through off-diagonal covariance terms. The correlations of controls introduced by observation uncertainty assimilation are found to dominate the reduction of uncertainty of transport. It is shown that coupling of a priori uncertainties that are assumed independent has the largest effect on uncertainty reduction. We note that while \citet{Cunningham2003} present detailed estimation error calculations, they assume all error components to be independent adding them as a quadrature sum, following eqn.~\ref{eq_quadrature_sum}. 
Here we demonstrate that cross-correlation between estimation error components modify significantly the resulting estimation uncertainty. 

The method provides a transparent framework for exposing the transformations of the assimilated uncertainty in terms of adjoint dynamics, relating uncertainty quantification to sensitivity analysis and highlighting the role of model sensitivities in analyzing numerical models. Adjoint barotropic waves are identified in the spatio-temporal uncertainty structures evolution, carrying information as uncertainty waves. Adjoint Kelvin and Rossby uncertainty waves are identified, carrying uncertainty information in the opposite (time-reversed) direction of the actual waves. 
Wave dynamics is contrasted with the leading order exponential dynamics of linear geostrophic adjustment, which explains the opposite uncertainty mechanisms of initial and boundary condition fields. Effects of initial condition uncertainty are transient and decay in the steady state, while the steady boundary conditions (here wind forcing) and numerical parameter (here dissipation) uncertainties grow from zero to steady state. 

Looking forward, our 
next goal is to integrate the developed machinery in the full realistic configuration of the ECCO ocean state estimation system. This will require the extension from synthetic to real observations, taking into account the heterogeneous data streams and associated spatio-temporal sampling characteristics, and to include the full three-dimensional baroclinic ocean dynamics. 
An added challenge will be the high dimensionality of the observation space.
The developed UQ method can be applied for model calibration to guide selection of physical parameterizations, boundary conditions, numerical parameters, not only in oceanography, but also in glaciological applications (e.g., \cite{Goldberg2015}). The uncertainty assimilation method is applicable to new observation systems design by quantifying the expected information gains and optimizing for the specific observation goals. 
Understanding the time dimension of uncertainty can be utilized for experiment design to account not only for geographic effects due to nonuniform uncertainty patterns but also for transient and steady consequences of the resulting inferences. 
Future research directions include extending the present 
 deterministic
framework to statistical inference in order to formally quantify coverage probabilities, goodness of fit and uncertainty partition between resolved and unresolved uncertainty sources. Improving our understanding and quantification of uncertainty dynamics of the real ocean constitutes an important contribution of computational science to the grand challenge problem of ocean and climate observability and predictability.


%


\acknowledgments
This work is supported in part by NSF “Collaboration in Mathematical Geosciences” (CMG) grant \#0934404 and DOE/SciDAC grant \#SC0008060 (PISCEES). It is a contribution to the overall ECCO effort, whose sustained support over the years through NASA is gratefully acknowledged.

 \bibliographystyle{ametsoc2014}
 \bibliography{references}

\end{document}